\documentclass[journal]{IEEEtran}

\usepackage[cmex10]{amsmath}
\usepackage{amscd,amsfonts,amsmath,amssymb,mathrsfs,pifont,stmaryrd,tipa}
\usepackage{array,cases,dsfont,graphicx,texdraw}
\usepackage{graphics} 
\usepackage{epsfig} 
\usepackage{subfigure}
\usepackage{algorithm,algorithmic}

\usepackage{colortbl}

\newtheorem{Remark}{Remark}
\newtheorem{Corollary}{Corollary}
\newtheorem{Definition}{Definition}
\newtheorem{Problem}{Problem}
\newenvironment{Proof}{\noindent{\em Proof:\/}}{\hfill $\Box$\par}
\newtheorem{Theorem}{Theorem}
\newtheorem{Lemma}{Lemma}

\newtheorem{Assumption}{Assumption}
\newtheorem{Proposition}{Proposition}

\newcommand{\EQQ}{\begin{eqnarray*}}
\newcommand{\ENN}{\end{eqnarray*}}
\newcommand{\EQ}{\begin{eqnarray}}
\newcommand{\EN}{\end{eqnarray}}

\renewcommand{\tt}{{\mbox{\tiny\sf T}}}
\renewcommand{\t}{{\mbox{\scriptsize \sf T}}}

\usepackage{lineno}
\ifCLASSINFOpdf
\else
\fi
\begin{document}
%
\title{Invariance Principles and Observability in Switched Systems with an Application in Consensus}%
%

\author{Ti-Chung~Lee,~\IEEEmembership{Senior Member,~IEEE}, Ying~Tan,~\IEEEmembership{Senior Member,~IEEE}, Youfeng~Su,~\IEEEmembership{Member,~IEEE}, and Iven~Mareels,~\IEEEmembership{Fellow,~IEEE}

\thanks{This work was supported in part by the MOST, Taiwan, R.O.C., under contract MOST 106-2221-E-159-001-MY2 and in part by National Natural Science Foundation of
China under Grant No. 61773122.} 

\thanks{T. C. Lee is with the Department of Electrical Engineering, Minghsin University of Science and Technology, Hsin-Hsing Road, Hsin-Fong, Hsinchu, Taiwan 304, R.O.C. Email: \texttt{tc1120@ms19.hinet.net}.}

\thanks{Y. Tan is with the Department of Electrical and Electronic Engineering, the University of Melbourne, VIC 3010, Australia. Email: \texttt{yingt@unimelb.edu.au}.}

\thanks{Y. Su is with the College of Mathematics and Computer Science, Fuzhou
University, Fuzhou 350116, P. R. China. Email: \texttt{yfsu@fzu.edu.cn}.}

\thanks{I. M. Mareels is with IBM Research Australia, 60 City Road, South Bank, VIC, 3006, Australia. Email: \texttt{imareels@au1.ibm.com}.}

}


\maketitle

\begin{abstract}
Using any nonnegative function with a nonpositive derivative along trajectories to define a virtual output, the classic LaSalle invariance principle can be extended to switched nonlinear time-varying (NLTV) systems, by considering the weak observability (WO) associated with this output.  WO is what the output informs about the limiting behavior of state trajectories (hidden in the zero locus of the output). In the context of switched NLTV systems, WO can be explored using the recently established framework of limiting zeroing-output solutions. Adding to this, an extension of the integral invariance principle for switched NLTV systems with a new method to guarantee uniform global attractivity of a closed set (without assuming uniform Lyapunov stability or dwell-time conditions) is proposed. By way of illustrating the proposed method, a leaderless consensus problem for nonholonomic mobile robots with a switching communication topology is addressed, yielding a new control strategy and a new convergence result.
\end{abstract}


\IEEEpeerreviewmaketitle

\section{Introduction} \label{sec-intro}

\IEEEPARstart{T}{he} classic LaSalle invariance principle \cite{LaS} has played an important role in checking attractivity of nonlinear \emph{time-invariant} systems. In its simplest form, the LaSalle invariance principle uses a sufficiently smooth function $V$ whose derivative along the trajectories is negative semi-definite. It then suffices to analyze the invariant set contained in the set of states corresponding to $\mathrm{d} V/ \mathrm{d} t=0$. As this nonlinear system is time-invariant, the convergence to the invariant set can be represented as the convergence of a single trajectory starting from the initial time instant $t_0=0$ (or attractivity only). The concept of the integral invariance principle was also proposed to relax the requirement that the derivative of $V$ has to be negative semi-definite \cite{BM}.

It is usually difficult to generalize the idea of the LaSalle invariance principle to \emph{time-varying} systems to conclude uniform attractivity due to the difficulty to characterize the invariant set when a family of trajectories from any initial time instant are considered. There are several attempts to extend this classic result to time-varying systems \cite{AM,ASP,Art,LLC,LJ2005,NA}. For example, the concept of limiting equations was proposed to capture the limiting behavior of the state trajectories \cite{Art, LLC}.  Moreover, the concept of virtual output was proposed for time-varying systems with the corresponding observability \cite{AM,ASP}.
 The advantage of introducing such a virtual output
is two-fold. Firstly, it is easier to check the
limiting behavior of signals rather than identifying a ``time-varying" (invariant) set. Secondly, several tools,  such as the persistent excitation condition
and observability/detectability,  are available to link the convergence of output signals  to the convergence of state trajectories.

Recently, switched systems have gained a lot of attention due to the possible performance improvement by introducing switching signals \cite{BRA, DBPL, JGG, LJ2008, Lib, OM}. Nevertheless, introducing switching in dynamical systems makes stability analysis much more challenging. Regarding the extension of the LaSalle invariance principle to switched systems, the majority of the literature assumes that for a given value of the switching signal, the ``present'' system behavior can be described by a time-invariant system, and moreover, that switching does not occur too often (so called dwell-time conditions are imposed). For example, linear time-invariant systems were considered in \cite{CWH, Hes} and nonlinear time-invariant systems were considered in \cite{BM2005,GST2008, MG2011, SGT2007}. When switching signals are arbitrary, only linear time-invariant systems were considered \cite{BJ, RSD}.

To the best of the authors' knowledge, there is no generalizations of either the use of virtual output or the LaSalle invariance principle to switched nonlinear time-varying (NLTV) systems when the switching signals do not have dwell-time constraints. Moreover, as pointed out in \cite{LTM2017}, the time-varying nature of nonlinear systems as well as time-varying components coming from a family of switching signals pose significant technical challenges. In particular, the notion of uniform convergence to a time-varying invariant set is not trivial.

It is noted that the classic LaSalle invariance principle has been widely used to check uniform attractivity when a weak Lyapunov function\footnote{A weak Lyapunov function is a positive definite function whose derivative along the trajectories of dynamics is negative semi-definite.} exists. Such a weak Lyapunov function guarantees uniform Lyapunov stability of dynamic systems. It is worthwhile to highlight that some early attempts \cite{LTM2017, LTM2019, MG2018} on a switched NLTV system followed this direction with the assumption that the switched NTLV system has a weak Lyapunov function, and gave the generalization of the so-called Krasovskii-LaSalle theorem, which requires stronger assumptions than that in the classical LaSalle invariance principle. For switched NLTV systems, such results can only guarantee the convergence of an equilibrium point (or a compact set). In some applications, as demonstrated in the consensus problem in this work, it is necessary to guarantee the convergence to a non-compact set when it is not feasible to assert (uniform) Lyapunov stability.

This paper focuses on direct extensions of the classic LaSalle invariance principle and the integral invariance principle to switched NLTV systems allowing for a general class of switching signals without any dwell-time constraints. Particularly, by utilizing the concept of virtual output and the corresponding observability condition as in \cite{AM, Art, LLC, NA}, the needed extensions naturally links to the recently developed framework in \cite{LTM2019}. This framework employs the concept of limiting zeroing output solutions, resulting in techniques such as changing state functions (dynamics) and output functions (output signals). These techniques provide great flexibility in checking the appropriate observability (see examples presented in \cite{LTM2019}).

In order to generalize this classic result without the existence of a weak Lyapunov function,  this paper introduces the concepts of uniformly globally ultimately bounded (UGUB) solutions and  weak observability (WO) for switched NLTV systems under a general class of switching signals. More precisely, this paper explores the concept of the virtual output to check uniform global attractivity (UGA) with respect to initial time instants and a large class of
switching signals when the switched NLTV system is already  UGUB. In the sequel, the techniques of  changing dynamics
and output signals presented in \cite{LTM2019} are modified accordingly. Moreover, the integral invariance principle can be extended to switched NLTV systems as well.

One of the significant practical motivations for this study is to establish efficient analysis tools for the consensus of multi-agent systems with switching topologies. In fact, the switching topology induces a switching controller, and hence results in a closed-loop system exhibiting switching dynamics. In general, the consensus is achieved if the so-called consensus subspace is attractive, which is non-trivial even when all agents are single integrator systems \cite{JLM2003,Lin2005,RB2005}. When the switching topology satisfies the so-called jointly connected condition (UJC), one can find a weak Lyapunov function, and hence, dwell-time conditions are frequently imposed to show the attractivity \cite{Lin2005,SH2012,SH2012b,YMSHJ2006}. Notice that a  dwell-time condition does not fit the practical circumstance of switching communication topologies, as one may not be able to associate a dwell-time to link failures. It is therefore important to consider approaches that avoid dwell-time constraints.

Regarding the consensus of nonholonomic mobile robots, most of studies either address the static network topologies or require that the switching network topologies are always represented as connected graphs, see \cite{DK2007,DF2008,LFM2005,MBNLP2019,YXLF2018}, to name just a few. Notice that, even under these simplifying conditions, due to Brockett's necessary condition, the time-varying feedback control is essential. That is why one has to deal with switched NLTV systems. One recent result in this direction can be found in \cite{LJ2014}. The authors do impose dwell-time conditions, and limited certain dynamics to obtain their results.

There
are as least three main challenges that one has to overcome to address attractivity in consensus. First, the closed-loop system is always a switched NLTV system, while the weak Lyapunov function is hard to find. Second, it is important to remove the dwell-time condition which is normally imposed in this literature \cite{JLM2003,Lin2005,LJ2014,RB2005,SH2012,SH2012b,YMSHJ2006}. This most likely indicates that a different analysis approach is necessary. Third, it is noticed that  each agent has three state variables (position and angle), but only has two control variables (velocity and angle velocity). Hence, it is necessary to inject certain persistently exciting signals in order to handle the variables that cannot be directly controlled \cite{LJ2005}, \cite{MBNLP2019}.

The proposed method is an ideal tool to solve such a challenge problem. That is, the proposed method will be used to develop consensus control laws for nonholomomic mobile robots without any dwell-time conditions. For this purpose, a generalized jointly connected condition will be provided that can be seen as the weakest condition since it is without the dwell-time condition and strictly includes standard UJC.  As will be shown in Section \ref{sec-4C} and summarized in Remark \ref{remark-11-0}, applying the proposed ``new invariance principle" for switched NLTV systems, the consensus can be achieved subject to a generalized jointly connected switching topologies without requiring any dwell-time conditions.  Moreover, in contrast with the previous consensus studies where only the attractivity is obtained, the presented design can further guarantee uniform global attractivity, giving rise to much more robustness.

The contributions of this paper are:

\begin{enumerate}
  \item  The verification of the convergence properties of the largest invariant set is converted into a  problem of checking WO employing a virtual output.

  \item By requiring uniformly globally ultimately bounded solutions, which is a much weaker property than uniformly globally bounded solutions, checking WO of a switched NLTV system becomes similar to checking weak detectability of this switched NLTV system \cite{LTM2019}. The concepts of changing dynamics of the state and changing output functions can be adapted to check WO of switched NLTV systems, providing more flexibility in verifying WO.

  \item The integral invariance principle is extended to switched NLTV systems.

  \item The proposed results are applied to leaderless consensus problems. To the best of the authors' knowledge, this is the first result and control design that achieves consensus without imposing a dwell-time condition.
  \end{enumerate}

This paper is organized as follows. Preliminaries and the problem formulation are presented in Section \ref{sec-pre}, followed by main results in Section \ref{sec-main}. Section \ref{sec-consensus} presents a new leaderless consensus problem to illustrate the effectiveness of the proposed methods. Conclusion is drawn in Section \ref{sec-conclusion}.

~

\textbf{Notations}
\begin{enumerate}
  \item  $\mathbb{N}=\{1,2,\dots\}$, $\mathbb{Z}=\{0,\pm1,\pm2,\dots\}$, $\mathbb{Z}_+=\{0,1,2,\dots\}$, $\mathbb{R}=(-\infty,+\infty)$, and $\mathbb{R}_{\geq 0}=[0,+\infty)$.

  \item  $\mathbb{R}^p$ denotes the $p$ dimensional Euclidean space and $\mathbb{R}^{p\times q}$ denotes the set of all $p\times q$ matrices with real entries.

  \item $\vert t\vert$ denotes the absolute value of a real number $t$ and $\Vert u \Vert$ denotes the Euclidean norm of a vector $u\in\mathbb{R}^{p}$.

  \item A function $g: \mathbb{R}_{\geq 0}\times \mathbb{X} \rightarrow \mathbb{R}^q$ with $\mathbb{X} \subseteq \mathbb{R}^p$ is said to be a Caratheodory function if for almost all $t\in\mathbb{R}_{\geq 0}$, $g(t,\cdot)$  is continuous and for each $u \in X$, $g(\cdot,u)$ is measurable.

  \item The function $g$ is said to be almost uniformly bounded if there is a measure zero set $\mathbb{E} \subseteq \mathbb{R}_{\geq 0}$  such that for any $r>0$,  there exists $M\triangleq M(r)>0$ satisfying $\Vert g(t,u)\Vert \leq M$, $\forall t\in\mathbb{R}_{\geq 0}\setminus \mathbb{E}$, $\forall u \in \mathbb{X}$ with $\Vert u \Vert\leq r$.

  \item The function $g$ is said to be continuous in $u$,  almost uniformly in $t$  if there exists a measure zero set $\mathbb{E} \subseteq \mathbb{R}_{\geq 0}$  such that for any $u \in \mathbb{X}$  and any $\varepsilon >0$,  there exists $\delta \triangleq \delta(\varepsilon,u)>0$  satisfying $\Vert g(t,u)-g(t,v)\Vert <\varepsilon$, $\forall t\in\mathbb{R}_{\geq 0}\setminus E$, $\forall v \in \mathbb{X}$ with $\Vert v-u\Vert < \delta$.

  \item For a set $\mathbb{X} \subseteq \mathbb{R}^p$ , the set $\mathrm{CB}(\mathbb{X})$ is the family of all functions $g: \mathbb{R}_{\geq 0}\times \mathbb{X} \rightarrow \mathbb{R}^q$  that are Caratheodory and almost uniformly bounded.

  \item For a set  $\mathbb{X} \subseteq \mathbb{R}^p$, the set $\mathrm{CC}(\mathbb{X})$  is the family of all functions $g(t,u) \in \mathrm{CB}(\mathbb{X})$  being continuous in $u$,  almost uniformly in $t$.

  \item For any $n \in \mathbb{N}$, $1_n=[{1,\dots,1}]^{\t}\in \mathbb{R}^n$.


  \item For any function $g:~\mathbb{X}\rightarrow \mathbb{R}^p$ and any subset $\hat{\mathbb{X}} \subseteq \mathbb{X}$, $\left.g\right|_{\hat{\mathbb{X}}}:~\hat{\mathbb{X}} \rightarrow \mathbb{R}^p$ denotes the restriction function of $g$, i.e., $\left.g\right|_{\hat{\mathbb{X}}}(u)=g(u)$, $\forall u \in \hat{\mathbb{X}}$.

  \item For any function $g: \mathbb{X} \rightarrow \mathbb{Y}$ and each $y\in \mathbb{Y}$, $g^{-1}(y)=\{x\in \mathbb{X}|~g(x)=y\}$.
\end{enumerate}

\section{Preliminaries}  \label{sec-pre}

Here, the LaSalle invariance principle is briefly reviewed and reformulated as an observability condition for a virtual output. Next appropriate stability definitions for NLTV systems are presented. The limiting zero output condition \cite{LTM2019} is recalled to introduce the concept of weak observability, which will be elaborated in the next section.

\subsection{From Invariance Principles to Attractivity} \label{sec-02A}

Recall the invariance principle for the following nonlinear continuous system:
\begin{align}
\dot{x} = f(x),~x \in \mathbb{R}^p. \label{eq-nonlearsystem}
\end{align}
Suppose there exists a continuously differentiable function
$V:~\mathbb{R}^p\rightarrow\mathbb{R}_{\geq 0}$ such that
\begin{align}\label{eq-dotV}
\dot{V} = \frac{\partial V(x)}{\partial x}f(x) \leq 0.
\end{align}
Then, the celebrated \emph{LaSalle invariance principle} guarantees that any bounded (forward complete) solution $x:~\mathbb{R}_{\geq 0} \rightarrow \mathbb{R}^{p}$ of the system \eqref{eq-nonlearsystem} approaches the largest invariant set contained in
$
\dot{V}^{-1}(0) = \{x \in \mathbb{R}^p|~\dot{V}(x)=0 \},
$
see \cite{Kha} for instance. Here the inequality \eqref{eq-dotV} can be relaxed to the following output converging condition
\begin{align}\label{eq-eb}
\int_{0}^{+\infty} \vert y(\tau)\vert^2 \mathrm{d} \tau \leq V(x(0)) < \infty
\end{align}
by defining a virtual output $y(t)= (-\dot{V}(x(t))^{1/2}$ for all $t \in \mathbb{R}_{\geq 0}$, see \cite{BM}, or even, by the Cauchy condition,
\begin{align}\label{eq-cauchy}
\lim_{t\rightarrow +\infty} \int_{t}^{t+T} \vert y(\tau)\vert^2 \mathrm{d} \tau = 0
\end{align}
for any (fixed) $T>0$ \cite{Lan}. This relaxation may accommodate a larger class of Lyapunov like functions whose derivatives along trajectories are nonpositive  at every instant of time.
Furthermore, let $\Omega \subseteq \mathbb{R}^p$ be the largest invariant set contained in $\dot{V}^{-1}(0)$.
Then, $x(t)\rightarrow \Omega$ as $t \rightarrow \infty$ describes the attractivity of   the closed set $\Omega$.
Moreover, the fact that $\Omega$ is the largest invariant set contained in $\dot{V}^{-1}(0)$ can be interpreted as the following observability condition:
\begin{align}\label{eq-obsp}
y(t)=0,~\forall t \in \mathbb{R}_{\geq 0} ~ \Longrightarrow ~x(t) \in\Omega,~\forall t \in \mathbb{R}_{\geq 0}.
\end{align}
Such a result is the so-called \emph{integral invariance principle} proposed in \cite{BM}, summarized in the below.

\begin{Proposition}\label{theorem-TI}
  Consider the system \eqref{eq-nonlearsystem} where $f:~\mathbb{R}^p\rightarrow\mathbb{R}^p$  is a continuous function. Let $h:~\mathbb{R}^p\rightarrow\mathbb{R}^q$ be a continuous function and $\Omega \subseteq \mathbb{R}^p$ a closed set. Suppose the observability condition \eqref{eq-obsp} holds where $y(t) = h(x(t))$ for all $t \in \mathbb{R}_{\geq 0}$.  Then, each bounded solution $x:~\mathbb{R}_{\geq 0}\rightarrow\mathbb{R}^p$   of \eqref{eq-nonlearsystem} with the property that \eqref{eq-cauchy} (or \eqref{eq-eb}) holds for some $T>0$  satisfies $x(t)\rightarrow \Omega$  as $t\rightarrow \infty$. 
\end{Proposition}

\begin{Remark}
The LaSalle invariance principle only requires $V$ to be continuously differentiable. $V$ need not be positive nor definite. The key feature exploited in the LaSalle invariance principle is the time-invariant nature of the nonlinear dynamics, which simplifies greatly the notion of invariant set. In the context of time-varying systems, the invariant set is a much more difficult concept, and this in turn makes it hard to extend the LaSalle invariance principle to time-varying systems \cite{Art, LLC}. The approach taken here is to examine the limiting behavior of the trajectories directly.
\end{Remark}

In this paper, such invariance principles are further generalized to switched NLTV systems. A new problem formulation and more general criterion to conclude uniform convergence properties are presented, with
not only global attractivity of a single trajectory but also uniform global attractivity of a family of solutions. For this purpose, two significant issues need to be addressed.
One is to specify a generalized output converging condition like \eqref{eq-cauchy} for switched NLTV systems.
The other is to describe a novel observability property similar to \eqref{eq-obsp} for switched NLTV systems, based on the concept of \emph{limiting zeroing-output solutions} given in \cite{LTM2019}.

\subsection{Switched NLTV Systems: Stability Concept} \label{sec-02B}

This paper  considers the switched NLTV system
\begin{subequations}\label{eq-system}
\begin{align}
\dot{x}&=f(t,x,\lambda)\label{eq-system-state}\\
y&=h(t,x,\lambda)\label{eq-system-output}
\end{align}
\end{subequations}
where $t \in \mathbb{R}_{\geq 0}$, $x \in \mathbb{R}^p$, $y \in \mathbb{R}^q$, and $\lambda$ is the $\Lambda$-valued switching signal with $\Lambda$ being a finite index set; the nonlinear mappings $f:~\mathbb{R}_{\geq 0}\times \chi \rightarrow \mathbb{R}^p$ and $h:~\mathbb{R}_{\geq 0}\times \chi \rightarrow \mathbb{R}^q$, with a nonempty set $\chi \subseteq \mathbb{R}^p\times \Lambda$, are the state and output functions, respectively. Here, following the notation in \cite{Hes, LJ2008}, a switching signal $\lambda$ is defined as a piecewise constant and right-continuous function with finitely many discontinuous points in any finite time interval.

Let $t_0\geq 0$ and $(x,\lambda)$ be any pair of signals with $x:~[t_0,+\infty)\rightarrow \mathbb{R}^p$  and $\lambda:~[t_0,+\infty)\rightarrow \Lambda$  being a switching signal. It is said that $(x,\lambda)$  is a forward complete solution pair of \eqref{eq-system-state} w.r.t. $\chi$  if $x$  is locally absolutely continuous, $(x(t),\lambda(t)) \in \chi$,  and the following equality holds:
\begin{align}
 x(t)=x(t_0)+\int_{t_0}^{t}f(\tau,x(\tau),\lambda(\tau))\mathrm{d}\tau,~\forall t \in [t_0,+\infty).
\end{align}
As in \cite{LTM2019}, the following notations are used.
\begin{itemize}
  \item $\mathrm{Ini}(x) = \mathrm{Ini}(\lambda) = t_0$ when $x$ and $\lambda$ are defined on $[t_0,+\infty)$.

  \item $\chi_\zeta = \{u \in \mathbb{R}^p|~(u,\zeta) \in \chi\}$ for all $ \zeta \in \Lambda $.

  \item $\Phi$ denotes a set of forward complete solution pairs and
  \begin{align*}
\Phi^{st} &= \{x|~\exists~\lambda~s.t.~(x,\lambda) \in \Phi\},\\
\Phi^{sw} &= \{\lambda|~\exists~x~s.t.~(x,\lambda) \in \Phi\}.
  \end{align*}

  \item For any function $g:~\mathbb{R}_{\geq 0} \times \chi \rightarrow \mathbb{R}^q$,
  \begin{align*}
g_\zeta(t,u) = g(t,u,\zeta),~\forall \zeta \in \Lambda, \, \forall t \in \mathbb{R}_{\geq 0}, \, \forall  u \in \chi_{\zeta}.
  \end{align*}
\end{itemize}
To guarantee the existence of solutions and ensure the measurability of the output function, it is always assumed that for each  $\zeta \in \Lambda$, $f_{\zeta}$  and $h_{\zeta}$ are \emph{Caratheodory} functions. We also assume that $\chi _\zeta $  is closed for any $\zeta\in \Lambda$.

In this paper, it is assumed that some forward complete solutions exist. We need to check the stability properties of these forward complete solutions. To this end, the following stability properties are adopted \cite{LJ2008}. 

\begin{Definition}\label{def-02}
   Let $\Phi$ be a set of some forward complete solution pairs  $(x,\lambda)$ of \eqref{eq-system-state}.
\begin{enumerate}
 \item  It is said that $\Phi^{st}$  is \emph{uniformly globally ultimately bounded (UGUB)}  if, for any $r>0$ there exist $T\triangleq T(r) >0$  and $M\triangleq M(r) >0$   such that for any $x\in \Phi^{st}$  and any pair $(s,t)$  of nonnegative real numbers with $\Vert x(s)\Vert \leq r$  and  $t\geq s+T \geq s \geq \mathrm{Ini}(x)$,  we have $\Vert x(t)\Vert \leq M$. If $T$  can be arbitrarily chosen, $\Phi^{st}$ is said to be \emph{uniformly globally bounded} (UGB).

\item It is said that the origin is \emph{uniformly Lyapunov stable} (ULS) w.r.t. $\Phi$  if, for any $\varepsilon>0$  there exists $\delta \triangleq \delta(\varepsilon)$  such that for any  $x\in \Phi^{st}$ and any pair $(s,t)$  of nonnegative real numbers with  $\Vert x(s)\Vert < \delta$  and  $t\geq s \geq \mathrm{Ini}(x)$,  we have $\Vert x(t)\Vert < \varepsilon$.

\item It is said that the origin is \emph{uniformly globally stable} (UGS) w.r.t. $\Phi$   when the origin is ULS and $\Phi^{st}$ is UGB.

\item It is said that the origin is \emph{uniformly globally attractive} (UGA) w.r.t. $\Phi$   if, for any  $r>0$ and any  $\varepsilon>0$ there exists $T\triangleq T(r,\varepsilon)$  such that for any $x\in \Phi^{st}$  and any pair $(s,t)$  of nonnegative real numbers with $\Vert x(s)\Vert \leq r$  and $t\geq s+T \geq s \geq \mathrm{Ini}(x)$,   we have $\Vert x(t)\Vert <\varepsilon$.

\item It is said that the origin is \emph{uniformly globally asymptotically stable} (UGAS) w.r.t.   $\Phi$ if it is UGS and UGA w.r.t.  $\Phi$.

\item It is said that the origin is \emph{uniformly globally exponentially stable} (UGES) w.r.t. $\Phi$  if, there exist $a>0$ and $b>0$ such that for any $x\in \Phi^{st}$, we have $\Vert x(t)\Vert \leq a e^{-b(t-s)}\Vert x(s)\Vert$ for all $t\geq s\geq \mathrm{Ini}(x)$.
 \end{enumerate}
 \end{Definition}

 To generalize the LaSalle invariance principle, the following special stability properties relative to a closed set are described, where the boundedness of solutions on some time intervals with respect to the Euclidean norm  is assumed.

\begin{Definition}\label{def-02}
   Let $\Omega\subseteq \mathbb{R}^p$ be  a closed set and $\Phi$ a set of some forward complete solution pairs $(x,\lambda)$ of \eqref{eq-system-state}.
\begin{enumerate}
 \item It is said that $\Omega$ is \emph{uniformly ultimately Lyapunov stable} (UULS) w.r.t. $\Phi$  and the Euclidean norm if, for  any  $r>0$ and any  $\varepsilon>0$ there exist $T\triangleq T(r,\varepsilon)$  and $\delta\triangleq \delta(r,\varepsilon)$   such that for any  $x\in \Phi^{st}$ and any pair $(s,t)$  of nonnegative real numbers with $t\geq s \geq T+\mathrm{Ini}(x)$, $\Vert x(\tau)\Vert \leq r$, $\forall \tau \geq s$, and $\Vert x(s)\Vert _{\Omega}< \delta$, we have $\Vert x(t)\Vert _{\Omega}< \varepsilon$. If $T$  can be arbitrarily chosen, it is said that the origin is \emph{uniformly Lyapunov stable} (ULS) w.r.t. $\Phi$  and the Euclidean norm.

 \item It is said that $\Omega$  is \emph{uniformly globally attractive} (UGA) w.r.t. $\Phi$  and the Euclidean norm if, for any $r>0$  and any $\varepsilon>0$  there exists $T\triangleq T(\varepsilon, r)>0$  such that for any $x\in \Phi^{st}$  and any pair $(s,t)$  of nonnegative real numbers with $\Vert x(s)\Vert \leq r$  and  $t\geq s+T \geq s \geq \mathrm{Ini}(x)$,  we have $\Vert x(t)\Vert _{\Omega}< \varepsilon$.
\end{enumerate}
 \end{Definition}

\begin{Remark}
  In the statement of the LaSalle invariance principle for switched NLTV systems, there exist two measures, namely, the Euclidean norm $\Vert \cdot\Vert $  and the $\Omega$-norm  $\Vert \cdot\Vert _{\Omega}$. The key assumption in the LaSalle invariance principle is the boundedness of solutions, which is measured in terms of Euclidean norm. However, the convergence property of the LaSalle invariance principle is relative to some set $\Omega$. For example, in the classic result for nonlinear time-invariant systems, this $\Omega$ is the largest invariant set. Therefore, two norms are used to characterize the stability properties.
\end{Remark}

\subsection{Limiting Zeroing-Output Solutions} \label{sec-02C}

The following definition was employed in \cite{LTM2019} in obtaining a generalized Krasovskii-LaSalle theorem for switched NLTV systems. The role of such a definition is to describe the convergence properties of output signals of \eqref{eq-system-state}. It will also play a key role in defining a suitable observability property in this paper.

\begin{Definition}\label{def-3}
A continuous function $\bar{x}:~\mathbb{R} \rightarrow \mathbb{R}^p$ is said to be \emph{a limiting zeroing-output solution} of the switched NLTV system \eqref{eq-system} w.r.t. $\Phi$ if, there exist sequences $\{(x_n,\lambda_n)\} \subseteq \Phi$  and $\{t_n\} \subseteq \mathbb{R}_{\geq 0}$  such that the following hold:
\begin{itemize}
  \item $t_n \geq \mathrm{Ini}(x_n) + 2n$.
  \item $\{x_n(\cdot+t_n):~[-n,n]\rightarrow \mathbb{R}^p\}$ converges uniformly to   $\bar{x}$   on every compact subset of $\mathbb{R}$.
  \item  For almost all $t \in \mathbb{R}$,
  \begin{align}
\lim_{n\rightarrow+\infty}  h (t+t_n, x_n(t+t_n), \lambda_n (t+t_n))=0.
  \end{align}
\end{itemize}
\end{Definition}

This is not a simple property, but observe that capturing the convergence of output signals in the context of switched NLTV systems is not trivial, and requires one to go back to the roots of limits through the consideration of (sub)sequences. Three sequences are used to reflect the nature of trajectories in limiting situation.  The first one is the initial time constant sequence $\{t_n\}$. The second one is the sequence $\{x_n\}$ of state signals. The last one is the sequence $\{\lambda_n\}$ of switching signals.

The concept of \emph{pairing} introduced in \cite{LTM2014} is also used in the sequel.

\begin{Definition}
For a nonempty set $\mathbb{X}$, let $g:~\mathbb{R}_{\geq 0} \times \mathbb{X} \rightarrow \mathbb{R}^q$  and $\hat{g}:~\mathbb{R}_{\geq 0} \times \mathbb{X} \rightarrow {\mathbb{R}^{\hat{q}}}$  be two functions. The pair $(g,\hat{g})$ is said to be \emph{a zeroing pair} if, for any time sequence $t_n\rightarrow + \infty$  and any sequence $\{u_n\} \subseteq \mathbb{X}$, the following holds:
\begin{align} \label{eq-zeroingpair}
\lim_{n\rightarrow+\infty}g(t_n,u_n) = 0 ~ \Longrightarrow ~\lim_{n\rightarrow+\infty}\hat{g}(t_n,u_n) = 0.
\end{align}
\end{Definition}

\begin{Remark}
 If two time-varying functions are a zeroing pair, they share a similar limiting behaviors though their transient behaviors might not be the same.
For example, $(e^{-t}+x, x)$ is a zeroing pair.
This concept will provide a flexibility in checking the limiting behaviors of switched NTLV systems.
\end{Remark}

\begin{Remark}
  Zeroing pairs can be easily found for time-invariant functions. Actually, it can be verified that if $\mathbb{X}$ is compact,  and $g_0:~\mathbb{X} \rightarrow \mathbb{R}^q$ and $\hat{g}_0:~\mathbb{X} \rightarrow \mathbb{R}^{\hat{q}}$ are time-invariant and continuous, then $(g_0,\hat{g}_0)$  is a zeroing pair if and only if for any $u \in X$, $g_0(u)=0$ $\Longrightarrow$ $\hat{g}_0(u)=0$.
\end{Remark}

Using the notion of zeroing pairs, one can  construct a switched system with new dynamics and new output instead of the original switched system \eqref{eq-system}, so that the limiting behaviors of the new switched NLTV system can represent the limiting behaviors of the original system when the following assumption holds:

\begin{Assumption}\label{ass-zeroingpair}
   There exist functions $\hat{f}:~\mathbb{R}_{\geq 0} \times \chi \rightarrow \mathbb{R}^p$ and $\hat{h}:~\mathbb{R}_{\geq 0} \times \chi \rightarrow \mathbb{R}^{\hat{q}}$ such that for any $\zeta\in \Lambda$,\,\, $\hat{f}_{\zeta},\,\hat{h}_{\zeta}\in \mathrm{CC}(\chi_\zeta)$, and for any $r>0$, $(h_{\zeta}|_{\hat{\chi}_{\zeta}^r},(f_{\zeta}-\hat{f}_{\zeta})|_{\hat{\chi}_{\zeta}^r})$    and  $(h_{\zeta}|_{\hat{\chi}_{\zeta}^r},\hat{h}_{\zeta}|_{\hat{\chi}_{\zeta}^r})$ are both zeroing pairs, where
   $
   \hat{\chi}_{\zeta}^r = \mathbb{R}_{\geq 0} \times \{u \in \chi_\zeta \,|~\Vert u\Vert \leq r\}
   $.
\end{Assumption}

The following proposition holds, for a proof see \cite[Theorem 2]{LTM2019}.

\begin{Proposition}\label{prop-1}
Consider the switched NLTV system \eqref{eq-system}, where $f_{\zeta}\in \mathrm{CB}(\chi_{\zeta})$, for all $\zeta \in \Lambda$.  Suppose Assumption \ref{ass-zeroingpair} holds. Let $\Phi$ be a set of some forward complete solution pairs   of  \eqref{eq-system-state}. Then, every bounded limiting zeroing-output solution $\bar{x}:~\mathbb{R} \rightarrow \mathbb{R}^p$ satisfies the following conditions where $\{t_n\} \subseteq \mathbb{R}_{\geq 0}$  and  $\{\lambda_n\}\subseteq \Phi^{sw}$ with $t_n \geq \mathrm{Ini}(\lambda_n)+2n$:
\begin{enumerate}
  \item [a)] Define $$\bar{\Lambda}(t)=\{\zeta \in \Lambda|~\zeta=\lambda_n(t+t_n)\mbox{~for infinitely many~} n\}.$$ Then, for any $t\in  {\mathbb{R}}, $
  \begin{align}
   \bar{x}(t) \in \bigcap_{\zeta \in \bar{\Lambda}(t)}\chi_{\zeta}.  \label{eq-12}
  \end{align}

  \item [b)] For all $t\in \mathbb{R}$,
    \begin{align}
&\bar{x}(t)= \bar{x}(0)  \nonumber \\
   &~~~~~~~ + \lim_{n \rightarrow +\infty}\int_{0}^{t}\hat{f}(\tau+t_n,\bar{x}(\tau),\lambda_n(\tau+t_n))\mathrm{d} \tau. \label{eq-13}
  \end{align}

  \item [c)] For almost all $t\in \mathbb{R}$,
      \begin{align}
\lim_{n\rightarrow +\infty}\hat{h}(t+t_n,\bar{x}(t),\lambda_n(t+t_n))=0. \label{eq-14}
  \end{align}
\end{enumerate}
\end{Proposition}

Proposition \ref{prop-1} indicates that under Assumption \ref{ass-zeroingpair}, a limiting zeroing-output solution of the system \eqref{eq-system} can be represented by a limiting zeroing-output solution of another system $(\hat f, \hat h)$. As such a technique generates a new system by changing $(f,h)$ to $(\hat f,\hat h)$, this technique is called \emph{changing dynamics and output}.

\begin{Remark}
  To capture the limiting behavior of a switched time-varying system, we need to take care of two ``time-varying components'', namely, the time variable $t$ and the switching signal $\lambda$  appeared in the system function $f(t,x,\lambda)$.  The idea behind Proposition \ref{prop-1} is to simplify the original system and output functions simultaneously under the zeroing (original) output condition. One of important cases is that the (changed) new system function $\hat{f}$  is independent to the switching signal, i.e., $\hat{f}_{\zeta}=f_c$ for all $\zeta\in \Lambda$.  In this case, $\dot{x}=f_c(t,x)$   is called a common zeroing-output system, and the check of detectability and observability becomes much easier because the effect of switching in the state equation \eqref{eq-system-state} has been removed as discussed in \cite{LTM2017}. In this paper, this idea is used to reach consensus for nonholonomic mobile robots.
 \end{Remark}

\begin{Remark}
In general, to generate new systems as described in Proposition \ref{prop-1}, two limiting processes are necessary to deal with two time-varying components. Observe that sequences $\{t_n\} \subseteq \mathbb{R}_{\geq 0}$  and $\lambda_n \subseteq \Phi^{sw}$  are involved in Proposition \ref{prop-1}. For the time sequence $\{t_n\}$, one may consider the so-called limiting functions
$$
\begin{bmatrix}
    \bar{f}_{\zeta}(t,u) \\
    \bar{h}_{\zeta}(t,u) \\
\end{bmatrix}=\lim_{n \rightarrow +\infty}\begin{bmatrix}
    \hat{f}_{\zeta}(t+t_n,u) \\
    \hat{h}_{\zeta}(t+t_n,u) \\
\end{bmatrix},~\forall u \in \chi_{\zeta}$$
for almost all $t \in \mathbb{R}$ to replace the function $\hat{f}_{\zeta}$ and $\hat{h}_{\zeta}$ in \eqref{eq-13} and \eqref{eq-14}, respectively \cite{LLC, LJ2005, LJ2008}. As for the sequence $\{\lambda_n\}$ of switching signals, the compactness property of the space  $L_{\infty}(\mathbb{R})$, which consists of all essentially bounded functions, can be employed to establish its weak* convergence \cite[Theorem 3.16]{Rud}. Replacing the original sequence by its subsequence, we conclude that  there exist (limiting) functions $\bar{\lambda}^{\zeta}:~\mathbb{R} \rightarrow [0,1],~\zeta \in \Lambda$, satisfying
$$\lim_{n\rightarrow +\infty}\int_{0}^{t}\lambda_{n}^{\zeta}(\tau+t_n)\gamma(\tau)\mathrm{d}\tau = \int_{0}^{t}\bar{\lambda}^{\zeta}(\tau)\gamma(\tau)\mathrm{d}\tau,~\forall t \in \mathbb{R}$$
for any Lebesgue integrable function $\gamma$, where for any switching signal $\lambda$, $\lambda^{\zeta}:~\mathbb{R}\rightarrow\{0,1\}$, $\zeta\in \Lambda$, are the indicator functions, i.e.,
\begin{align}\label{eq-add-001}\lambda^{\zeta}(\tau) = \left\{
                           \begin{array}{c}
                           1,\mbox{~if~}\lambda(\tau)=\zeta \\
                           0,\mbox{~if~}\lambda(\tau)\neq\zeta \\
                           \end{array}
                         \right..
\end{align}
Notice that the limiting function $\bar{\lambda}^{\zeta}$, $\zeta \in \Lambda$, may not be a proper switching signal of itself. However, it has the property that $\sum_{\zeta \in \Lambda}\bar{\lambda}^{\zeta}(t)=1$ for all $t \in \mathbb{R}$.  Now observe that
$$
\hat{f}(t,u,\lambda(t))=\sum_{\zeta\in \Lambda}\lambda^{\zeta}(t)\hat{f}_{\zeta}(t,u),~\forall t\in \mathbb{R}_{\geq 0},~\forall u \in \chi_{\lambda(t)}.
$$
By taking the limit, \eqref{eq-13} becomes
\begin{align*}
\bar{x}(t)&=\bar{x}(0)+\lim_{n\rightarrow +\infty}\int_{0}^{t} \sum_{\zeta \in \Lambda} {\lambda}^{\zeta}_{n}(\tau+t_n) \hat{f}_{\zeta}(\tau+t_n,\bar{x}(\tau)) \mathrm{d} \tau\\
&=\bar{x}(0)+\lim_{n\rightarrow \infty} \int_{0}^{t}  \sum_{\zeta \in \Lambda}{\lambda}^{\zeta}_n(\tau+t_n) \bar{f}_{\zeta}(\tau,\bar{x}(\tau)) \mathrm{d} \tau\\
&=\bar{x}(0)+  \int_{0}^{t}  \sum_{\zeta \in \Lambda} \bar{\lambda}^{\zeta}(\tau) \bar{f}_{\zeta}(\tau,\bar{x}(\tau)) \mathrm{d} \tau.
\end{align*}
That is to say that $\bar{x}$  is a solution of $\dot{\bar{x}}=\sum_{\zeta \in \Lambda}\bar{\lambda}^{\zeta}(t)\bar{f}_{\zeta}(t,\bar{x})$.  Similarly, \eqref{eq-14} can be replaced by $\sum_{\zeta \in \Lambda} \bar{\lambda}^{\zeta}(t)\bar{h}_{\zeta}(t,\bar{x}(t))=0$ for almost all $t\in \mathbb{R}$, see \cite{BJ, MG2018, RSD} for some discussion on this.
\end{Remark}


\section{A Generalized Invariance Principle for Switched NLTV Systems} \label{sec-main}

In this section, the concept of limiting zeroing-output solution is used to define an observability property relative to a closed set $\Omega$ for switched NLTV systems. Then, a generalized invariance principle is developed to guarantee uniform global attractivity of $\Omega$, with three special cases to relax the uniform convergence of output signals. The proofs for the main results are presented in the Appendices.

\subsection{Weak Observability and Generalized Invariance Principle} \label{sec-03A}

\begin{Definition}\label{Def-5}
It is said that $\Omega$ is \emph{weakly observable} (WO) w.r.t. $\Phi$ if, every bounded limiting zeroing-output solution $\bar{x}:~\mathbb{R} \rightarrow \mathbb{R}^p$  of the system \eqref{eq-system} w.r.t. $\Phi$  satisfies $\bar{x}(0) \in \Omega$.
\end{Definition}

To guarantee the convergence behavior of the switched NLTV system \eqref{eq-system-state}, the following generalized convergence condition is needed. Roughly speaking, it says that if $x$ is bounded, and bounded away from $\Omega$ at some sufficiently large time instant $s'$, then there exists another sufficiently large time instant $s$ such that  $x$ is also bounded away from $\Omega$ at this time instant $s$ and yields a sufficiently small output signal on a sufficiently large time interval centered at $s$, simultaneously. This, together with WO, implies UGA of $\Omega$ as shown in the below Theorem \ref{theorem-main} (see Appendix \ref{sec-05A} for the proof).

\begin{Assumption}\label{ass-g-convergence}
 For any $\varepsilon >0$ and any $r>0$, there exist $T\triangleq T(\varepsilon,r)>0$, $\delta\triangleq \delta(\varepsilon,r)>0$, and a function $\alpha:~\mathbb{N}  \rightarrow\mathbb{N}$, which may depend on $\varepsilon$  and $r$, such that for any $n\in \mathbb{N}$ and any $t\geq T$,  if $(x,\lambda) \in \Phi$  with
 \begin{itemize}
   \item $\Vert x(\tau)\Vert \leq r$, for all $\tau \geq t+ \mathrm{Ini}(x),$ and
   \item $\Vert x(s')\Vert _{\Omega}\geq \varepsilon ~\mbox{for some}~s' \geq \alpha(n)+t+\mathrm{Ini}(x),$
 \end{itemize}
 we have $\Vert x(s)\Vert _{\Omega}\geq \delta$  and
\begin{align}\label{eq-15}
\int_{s-n}^{s+n}\Vert h(\tau,x(t),\lambda(t))\Vert ^2 \mathrm{d} \tau \leq \frac{1}{n},
 \end{align}
for some $s\geq 2n+t+\mathrm{Ini}(x)$.
 \end{Assumption}

\begin{Theorem}\label{theorem-main}
  Consider a closed set $\Omega \subseteq \mathbb{R}^p$  and the switched NLTV system \eqref{eq-system} where $f_{\zeta}\in \mathrm{CB}(\chi_{\zeta})$ for all $\zeta \in \Lambda$.  Let $\Phi$ denote a set of some forward complete solution pairs $(x,\lambda)$  of \eqref{eq-system-state}. Suppose Assumption \ref{ass-g-convergence} holds, $\Omega$  is WO w.r.t. $\Phi$, and  $\Phi^{st}$ is UGUB. Then, $\Omega$   is UGA w.r.t. $\Phi$  and the Euclidean norm.
\end{Theorem}

\subsection{Some Special Cases} \label{sec-03B}

In Theorem \ref{theorem-main}, the uniform convergence property of output signals are characterized by Assumption \ref{ass-g-convergence}, which is very general. Sometimes, stronger convergence conditions of the output signals do exist in applications.  This subsection provides several simplified results of Theorem \ref{theorem-main} when Assumption \ref{ass-g-convergence} is replaced by other stronger but simpler output convergence assumptions.

\begin{Assumption}\label{ass-3}
   There exists $T_0>0$  such that the following equality holds:
\begin{align}\label{eq-convergent-sigle}
\lim_{t\rightarrow +\infty} \int_{t}^{t+T_0}\Vert h(\tau,x(\tau),\lambda(\tau))\Vert ^2\mathrm{d}\tau = 0.
\end{align}
\end{Assumption}

Obviously, Assumption \ref{ass-3} is much simpler than Assumption \ref{ass-g-convergence}. It is more like the concept of single trajectory used in the classic LaSalle invariant principle, which satisfies Assumption \ref{ass-3}.  This assumption leads to a convergence property with respect to the set $\Omega$ as stated in Corollary \ref{corollary-single} (see the proof in Appendix \ref{sec-05B}).

\begin{Corollary}\label{corollary-single}
    Consider a closed set $\Omega\subseteq \mathbb{R}^p$  and the switched system \eqref{eq-system} where  $f_{\zeta}\in \mathrm{CB}(\chi_{\zeta})$ for all $\zeta \in \Lambda$.  Let $(x,\lambda)$  be a forward complete solution pair of \eqref{eq-system-state}. Suppose Assumption \ref{ass-3} holds, $\Omega$  is WO w.r.t.  $\{(x,\lambda)\}$, and $x$ is bounded. Then, $\lim_{t\rightarrow +\infty}\Vert x(t)\Vert _{\Omega}=0$, i.e.,
    $x(t)\rightarrow \Omega$  as $t\rightarrow \infty$.
\end{Corollary}

\begin{Remark}
Here, we would like to mention that Corollary \ref{corollary-single} (and hence Theorem \ref{theorem-main}) can be viewed as a generalized LaSalle invariance principle (and an integral invariance principle) in switched NLTV systems. Actually, consider the special case that $\chi  = \mathbb{R}^ p \times \Lambda$ with $\Lambda$ being a singleton (without switching). Suppose $f$  and $h$  are both time-invariant and continuous. By choosing $\hat{f} = f$  and $\hat{h} = h$,  the limiting integral equation \eqref{eq-13} in Section \ref{sec-02C} is equivalent to saying that $\dot{\bar{x}}=f(\bar{x})$ and the zeroing-output equation \eqref{eq-14} is reduced to $h(\bar{x}(t))=0$  for almost all $t$ in $\mathbb{R}$.  Since $h$ is continuous, $h(\bar{x}(t))=0$ for all $t \in \mathbb{R}$ \cite{Lan}. The observability property stated in Section \ref{sec-02A} (see \eqref{eq-obsp}) then implies weak observability. Thus, the integral invariance principle (Proposition \ref{theorem-TI}) follows from Corollary \ref{corollary-single}.
\end{Remark}

Another special case of Assumption \ref{ass-g-convergence} is represented by Assumption \ref{ass-4} under UULS w.r.t. $\Phi$ and Euclidean norm, which is a slightly modified version of \cite[Assumption 1]{LTM2019}.

\begin{Assumption}\label{ass-4}
For any  $0< \varepsilon < 1$, there exist $T\triangleq T(\varepsilon) >0$ and $M\triangleq M(\varepsilon) >0$  such that for any $(x,\lambda) \in \Phi$  and any $(s,t)$  with $\mathrm{Ini}(x)+T \leq s \leq t$ and $\Vert x(\tau)\Vert \leq {1}/{\varepsilon}$, $\forall s\leq \tau \leq t$, the following integral inequality holds:
\begin{align}\label{eq-17}
\int_{s}^{t} \Vert h(\tau,x(\tau),\lambda(\tau))\Vert ^2 \mathrm{d} \tau \leq M+\varepsilon(t-s).
\end{align}
\end{Assumption}

Assumption \ref{ass-4} itself is not sufficient to guarantee Assumption \ref{ass-g-convergence}. However, for the closed set $\Omega$ that is UULS w.r.t. $\Phi$  and the Euclidean norm, Assumption \ref{ass-4} does imply Assumption \ref{ass-g-convergence}, and hence leads to the following corollary. Its proof is presented in Appendix \ref{sec-05C}.

\begin{Corollary}\label{corollary-family}
  Consider a closed set $\Omega \subseteq \mathbb{R}^p$  and the switched NLTV system \eqref{eq-system} where $f_{\zeta}\in \mathrm{CB}(\chi_{\zeta})$ for all $\zeta \in \Lambda$.  Let $\Phi$ denote a set of some forward complete solution pairs $(x,\lambda)$  of \eqref{eq-system-state}. Suppose Assumption \ref{ass-4} holds,
  $\Omega$ is WO w.r.t. $\Phi$, and $\Phi^{st}$ is UGUB. If $\Omega$ is UULS w.r.t. $\Phi$ and the Euclidean norm, then it is UGA w.r.t. $\Phi$  and the Euclidean norm.
\end{Corollary}

In practice, to verify Assumptions \ref{ass-3} and \ref{ass-4},
one may resort to the following simple and useful bounded-output-energy condition:

\begin{Assumption}\label{ass-5}
The following integral inequality holds:
\begin{align}\label{eq-add-01++}
\int_{s}^{+\infty} \Vert h(\tau,x(\tau),\lambda(\tau))\Vert ^2 \mathrm{d} \tau \leq \mu(x(s)),~\forall s \geq \mathrm{Ini}(x)
\end{align}
for some continuous function $\mu:~\mathbb{R}^p \rightarrow \mathbb{R}_{\geq 0}$.
\end{Assumption}

Obviously, \eqref{eq-add-01++} implies \eqref{eq-convergent-sigle} because the Cauchy condition holds. We can also show that \eqref{eq-add-01++} implies \eqref{eq-17}. In fact,
For any $0< \varepsilon < 1$, let $T>0$ be any given and $M=\max_{\Vert u\Vert \leq {1}/{\varepsilon}}\mu(u)+1$. Then, $0<M<+\infty$. Thus, for any $(x,\lambda) \in \Phi$  and any $(s,t)$  with $\mathrm{Ini}(x)+T \leq s \leq t$ and $\Vert x(\tau)\Vert \leq{1}/{\varepsilon}$, $\forall s\leq \tau \leq t$,
\begin{align*}
\int_{s}^{t} \Vert h(\tau,x(\tau),\lambda(\tau))\Vert ^2 \mathrm{d} \tau  &\leq\int_{s}^{+\infty} \Vert h(\tau,x(\tau),\lambda(\tau))\Vert ^2 \mathrm{d} \tau \\
&\leq \mu(x(s)) \leq M.
\end{align*}
Thus, \eqref{eq-17} holds. Therefore, according to Corollaries \ref{corollary-single} and \ref{corollary-family}, the following  can be deduced.

\begin{Proposition}\label{corollary-useful}
Consider a closed set $\Omega \subseteq \mathbb{R}^p$  and the switched NLTV system \eqref{eq-system} where $f_{\zeta}\in \mathrm{CB}(\chi_{\zeta})$ for all $\zeta \in \Lambda$.  Let $\Phi$ denote a set of some forward complete solution pairs $(x,\lambda)$  of \eqref{eq-system-state}. Suppose Assumption \ref{ass-5} holds. When $\Omega$ is WO w.r.t. $\Phi$, and $\Phi^{st}$ is UGB, $x(t)\rightarrow \Omega$ as $t\rightarrow \infty$ for any $x\in \Phi^{st}$. In addition, if $\Omega$ is ULS w.r.t. $\Phi$ and the Euclidean norm, then it  is UGA w.r.t. $\Phi$  and the Euclidean norm.
\end{Proposition}

\begin{Remark}
 As discussed in Section \ref{sec-02A}, \eqref{eq-add-01++} in Assumption \ref{ass-5} usually comes from  the existence of a weak Lyapunov function $V$.
This leads to a virtual output chosen as $h=(-\dot{V})^{1/2}$. To guarantee the attractivity of bounded solutions to a closed set $\Omega$, it remains to check weak observability of $\Omega$ \cite{LJ2005,LJ2008,LTM2017,LTM2019}.
\end{Remark}

In general, the set $\Omega$ can contain equilibria and limit cycles. Moreover, all the results (Theorem \ref{theorem-main}, and Corollaries \ref{corollary-single} and \ref{corollary-family}) can be re-stated when the set $\Omega$ only contains the origin. In particular, as a special case, $\Omega = \{ 0 \}$ will induce UGAS of the origin, giving rise to the following corollary.

\begin{Corollary}\label{corollary-family-0}
  Consider the switched NLTV system \eqref{eq-system} where $f_{\zeta}\in \mathrm{CB}(\chi_{\zeta})$ for all $\zeta \in \Lambda$.  Let $\Phi$ denote a set of some forward complete solution pairs $(x,\lambda)$  of \eqref{eq-system-state}. Suppose Assumption \ref{ass-5} holds and the origin is WO w.r.t. $\Phi$. If the origin is UGS w.r.t. $\Phi$, then it is UGAS w.r.t. $\Phi$.
\end{Corollary}

\section{Application to Consensus of Nonholonomic Mobile Robots}  \label{sec-consensus}

By way of example, this section shows how to use the results to study the consensus of nonholonomic mobile robots subject to jointly connected network topologies.
Moreover, we can use the tools constructively to develop a distributed control law to enforce consensus without requiring dwell-time conditions for the switching network graphs.

\subsection{Problem Statement}\label{sec-4A}

Consider a group of $N$ nonholonomic mobile robots modelled in the following simplified form, see \cite{MS1993},
\begin{align}
  \dot{x}_i&=\cos(\theta_i)v_i, \nonumber\\
 \dot{y}_i& = \sin(\theta_i)v_i, \nonumber\\
 \dot{\theta}_i &=w_i,~ 1 \leq i \leq N\label{eq-robot}
\end{align}
where $[x_i,y_i]^{\t}\in \mathbb{R}^2$ represents the position of the $i$-th agent on a fixed frame, $\theta_i\in \mathbb{R}$ represents the pose (i.e., the angle from the horizontal axis) of the $i$-th agent, and $[v_i,w_i]^{\t}\in \mathbb{R}^2$ represent the forward and angular velocities respectively, which can be viewed as the control input of the $i$-th agent.

The network topology among these $N$ agents is modeled by a switching undirect graph
$\mathcal{G}(\zeta)=(\mathcal {V},\mathcal {E}(\zeta))$, where $\mathcal {V}=\{1,\dots,N\}$ represents the $N$ agents, $\mathcal {E}(\zeta)\subseteq \mathcal {V} \times \mathcal {V}$, and $\zeta \in \Lambda$ with $\Lambda$ representing all possible graphs which are not necessary to be connected. For a $\Lambda$-valued switching signal $\lambda: \mathbb{R}_{\geq 0} \rightarrow \Lambda$, $(j, i) \in \mathcal {E}(\lambda (t))$ if and only if the control of the $i$-th agent can make use of the information of the $j$-th agent for feedback at the time instant $t$.

For any $\tau_a>0$ and any switching signal $\lambda$,
the $\tau_a$-joint graph over a time interval $[t_1,t_2)$ is defined as
\begin{align} \label{eq-35a}
\mathcal{G}^{\tau_a}_{\lambda}([t_1,t_2)) = \left(\mathcal {V},\bigcup_{\zeta \in \lambda_{\tau_a} [t_1,t_2)}\mathcal {E}(\zeta)\right)
\end{align}
where \begin{align} \label{eq-35a2}\lambda_{\tau_a} [t_1,t_2)=\left\{\zeta\in \Lambda\left|~\int_{t_1}^{t_2}\lambda^{\zeta}(\tau) \mathrm{d} \tau \geq \tau_a\right.\right\}\end{align}
with $\lambda^{\zeta}(\cdot)$ being the indicator function as in \eqref{eq-add-001}. In this paper, the switching graph $\mathcal{G}(\lambda (t))$ is assumed to satisfy the following generalized uniformly jointly connected (GUJC) condition.

\begin{Assumption}\label{ass-ujc}
  There exists a constant pair $(\tau_a,T)$ with $T\geq \tau_a >0$ such that for any $\lambda\in \Phi^{sw}$ and any $t \in \mathbb{R}_{\geq 0}$, the $\tau_a$-joint graph $\mathcal{G}^{\tau_a}_{\lambda}([t,t+T)) $ is connected for any $t \in \mathbb{R}_{\geq 0}$, in the sense that it contains a spanning tree.
\end{Assumption}

The leaderless consensus problem can be formulated as follows, where, without loss of generality, the initial time of the state and the switching signal is always assumed to be zero.

\begin{Problem}[Leaderless Consensus Problem]\label{pro-01}
Consider the system \eqref{eq-robot}. Find a distributed controller such that for any initial states $x_i(0),y_i(0),\theta_i(0)$, $i=1,\dots,N$, all solutions of the closed-loop system exist and are bounded for any $t \in \mathbb{R}_{\geq 0}$ with
\begin{align}\label{eq-consensus}
\lim_{t\rightarrow\infty}\left([x_i(t),
                                   y_i(t),
                                   \theta_i(t)]
-
[x_j(t),
y_j(t),
\theta_j(t)]
\right) =0
\end{align}
for all $i\neq j$, $i,j=1,\dots,N$.
 \end{Problem}

\begin{Remark}
The integral $\int_{t_1}^{t_2}\lambda^{\zeta}(\tau) \mathrm{d} \tau $ represents the accumulative length of all subintervals of $[t_1,t_2)$, in which $\lambda(t)=\zeta$. Actually,   given any time interval $[t_1,t_2)$, denote by $t_{k_1},t_{k_2},\dots,t_{k_l}$ the switching time instants in this interval, and $t_{k_0}\triangleq t_1$, $t_{k_{l+1}}\triangleq t_2 $. Then,
\begin{align}\label{eq-add-t1t2}
\int_{t_1}^{t_2}\lambda^{\zeta}(\tau) \mathrm{d} \tau = \sum_{i=0,\lambda(t_{k_{i}})=\zeta}^{l}(t_{k_{i+1}}-t_{k_{i}}).
\end{align}
Thus, \eqref{eq-35a} means that, for any switching mode $\zeta$, the accumulative length of all subintervals of $[t_1,t_2)$ in which $\lambda(t)=\zeta$ should be lower bounded by some positive constant $\tau_a$.
\end{Remark}

\begin{Remark}\label{remark-07-0}
It is of interest to relate Assumption \ref{ass-ujc} to the standard dwell-time condition, which
together with uniformly jointly connected (UJC) condition is frequently used in the recent literature, see \cite{JLM2003,Lin2005,LJ2014,RB2005,SH2012,SH2012b,YMSHJ2006}, to name just a few.
The dwell-time condition requires that each admissible switching signal $\zeta \in \Lambda$ has $s_{i+1}-s_{i}\geq \tau_d$ for some fixed constant $\tau_d$, where $s_i$ and $s_{i+1}$ are two sequenced switching time instants of $\lambda(t)$.
According to \eqref{eq-add-t1t2}, if $t_2-t_1\geq 2 \tau_d$ and $\lambda(t)=\zeta$ for some $t_1+\tau_d \leq t \leq  t_2-\tau_d$, then $\zeta \in \lambda_{\tau_d} [t_1,t_2)$. Notice that, under the UJC condition, there exists $T'>0$ such that the joint graph $G[t+\tau_d,t+\tau_d+T')$ defined as $(\mathcal {V},\cup_{t\in [t+\tau_d,t+\tau_d+T')}\mathcal {E}(\lambda (t)))$
is connected for any $t \in \mathbb{R}_{\geq 0}$. By the previous discussion,
$G_\lambda^{\tau_d} [t,t+2\tau_d+T')$ contains $G[t+\tau_d,t+\tau_d+T')$, and hence
$G_\lambda^{\tau_d} [t,t+2\tau_d+T')$ is also connected.
This means the GUJC condition is weaker than the UJC condition because no dwell-time conditions are assumed in the GUJC condition, see also the simulation result in Section \ref{sec-4D} for such a situation.
\end{Remark}

\begin{Remark}
 It is of interest to mention that some other practical cooperative control problems can be directly converted to Problem \ref{pro-01}. For example, replacing \eqref{eq-consensus} with the objective
\begin{align*}
\lim_{t\rightarrow\infty}\left(\begin{bmatrix}
                                   x_i(t) \\
                                   y_i(t) \\
                                   \theta_i(t)\\
\end{bmatrix} -
\begin{bmatrix}
x_j(t) \\
y_j(t) \\
\theta_j(t)\\
\end{bmatrix}\right) = \begin{bmatrix}
d_{xi}-d_{xj} \\
d_{yi}-d_{yj}\\
0\\
\end{bmatrix}
\end{align*}
for some pre-desired targets $[d_{xi},d_{yi}]^{\t}\in \mathbb{R}^2$, $i=1,\dots,N$, leads to the so-called formation control problem for nonholonomic mobile robots \eqref{eq-robot}, see  \cite{LFM2005}, \cite{MBNLP2019}, \cite{YXLF2018}. Such a problem can be easily converted to Problem \ref{pro-01} with new state valuables $[x'_{i},y'_{i},\theta'_i]^{\t}$ where $x_i'=x_i-d_{xi}$, $y_i'=y_i-d_{yi}$, and $\theta'_i=\theta_i$.
\end{Remark}

\subsection{Controller Design}\label{sec-4A0}

Choose $T_0 > 2T$ and a continuous function $c:~[0, - T + T_0/2 )\rightarrow \mathbb{R}$ with $
 \int_{0}^{- T + {T_0}/{2}} c(\tau) \mathrm{d} \tau \neq k\pi$, $k\in \mathbb{Z}
$,  where $T$ is the constant given in Assumption \ref{ass-ujc}. Let $p: \mathbb{R}\rightarrow \mathbb{R}$ be a piecewise continuous periodic function
with the period $T_0$, for which on the interval $[0,T_0)$,
\begin{align}\label{eq-62}
p(t)=\left\{\begin{array}{ll}
       0, & t\in[0,T), \\
       c(t-T), & t\in[T,T_0/2),  \\
       0, & t\in[T_0/2,T + T_0/2),  \\
       -c(t-T-T_0/2), & t\in[T+T_0/2,T_0).
     \end{array}\right.
\end{align}
Notice that $p(t)$ has an interesting property that
\begin{align}\label{eq-30}
\int_{0}^{T_0}p(\tau) \mathrm{d}\tau =0.
\end{align}

\begin{Remark}
 Though $T_0$ is assumed to be strictly larger than $2T$, it does not mean that
 $T$ should be known exactly. In practice, one can use an open loop estimation for getting a sufficiently large upper bound $T'>T$. Then, $T_0\geq 2 T'$ will be the one we need. For the special case that $\mathcal{G}(\zeta)$ is connected for all $\zeta \in \Lambda$, $T'$ can be chosen arbitrarily small, so is $T_0$.
\end{Remark}

For any $\zeta\in \Lambda$, let $\mathcal
{L}(\zeta)=[l_{ij}(\zeta)]\in \mathbb{R}^{N\times N}$ be the Laplacian matrix of the graph $\mathcal{G}(\zeta)$, where $l_{ij}(\zeta)=-a_{ij}(\zeta)$, $i\neq j$, and $l_{ii}(\zeta)=\sum_{j=1}^Na_{ij}(\zeta)$. Here for all $\zeta \in \Lambda$, $a_{ij}(\zeta)$ satisfy
$a_{ii}(\zeta)=0$, and $a_{ij}(\zeta)=a_{ji}(\zeta)>0$ if and only if $(j,i)\in \mathcal {E}(\zeta)$. Notice that $\mathcal
{L}(\zeta)$ is symmetric and positive semi-definite for all $\zeta \in \Lambda$ \cite{JLM2003,OM,RB2005}.

Then, a distributed controller can be designed as follows:
\begin{subequations}\label{control}
\begin{align}
v_i&=-k_{v}\sum_{j=1}^{N}a_{ij}(\lambda(t))\begin{bmatrix}\cos(\theta_i) & \sin(\theta_i)\end{bmatrix}\begin{bmatrix} x_i-x_j \\ y_i-y_j\end{bmatrix}, \label{control-v}\\
w_i&=p(t)-k_w\sum_{j=1}^{N}a_{ij}(\lambda(t))(\theta_i-\theta_j)\label{control-w}
\end{align}
\end{subequations}
where $k_v,k_w>0$ can be chosen arbitrarily.
With this distributed controller \eqref{control}, Problem \ref{pro-01} can be solved as stated in the following theorem.

\begin{Theorem}\label{Theorem-consensus}
  Under Assumption \ref{ass-ujc}, the distributed controller \eqref{control} solves Problem \ref{pro-01}.
\end{Theorem}

The proof will be given in the following two subsections. Specifically, in Section \ref{sec-4B}, the consensusability of the system \eqref{eq-robot} is
converted into the stability analysis of a closed set that is called the consensus subspace. Then, uniform global attractivity of the consensus subspace is guaranteed employing Proposition \ref{corollary-useful} based on the concept of weak observability.

\begin{Remark}
Like \cite{MBNLP2019}, in \eqref{control}, the angle information of each agent is assumed to be known by itself. In practice, the angle information can be measured by using the gyroscope.
\end{Remark}

\begin{Remark}
 Notice that the GUJC switching network topologies strictly include the UJC switching network topologies, hence Assumption \ref{ass-ujc} includes also the static network topologies and the switching network topologies with every switching graph being connected.
\end{Remark}

\subsection{Coordinate Transformation} \label{sec-4B}

To depict the closed-loop system, a new matrix is needed based on the graph Lapalacian matrix $\mathcal
{L}(\lambda(t))$. Given any $\zeta \in \Lambda$ and any $B=[b_{ij}] \in \mathbb{R}^{N \times N}$, let $\mathcal
{L}(\zeta,B)\in \mathbb{R}^{N \times N}$ have the $(i,i)$ entry being $\sum_{j=1}^{N}a_{ij}(\zeta)b_{ij}$ and the $(i,j)$, $i \neq j$, entry being $-a_{ij}(\zeta)b_{ij}$. Regarding this matrix, it holds that $\mathcal
{L}(\zeta)=\mathcal
{L}(\zeta,1_N 1_N^{\t})$.
Moreover, let $L_0=I-({1}/{N})1_N1_N^{\t}$, which can be viewed as a special case of $\mathcal {L}(\zeta)$  with $a_{ij}(\zeta) =  {1}/{N}$ for all $i \neq j$.

In the body frame,  consider the following transformation
\begin{align}\label{eq-transformation01}
\begin{bmatrix}
     {\tilde{x}}_i \\
     {\tilde{y}}_i \\
\end{bmatrix} &=\begin{bmatrix}
             \cos(\theta_i) & \sin(\theta_i) \\
             -\sin(\theta_i) & \cos(\theta_i)  \\
\end{bmatrix} \begin{bmatrix}
     {x}_{i} \\
     {y}_{i} \\
\end{bmatrix} .
\end{align}
Let ${x}=[\tilde{x}_1,\dots,\tilde{x}_N]^{\t}$, ${y}=[\tilde{y}_1,\dots,\tilde{y}_N]^{\t}$, ${\theta}=[{\theta}_1,\dots,{\theta}_N]^{\t}$, and $B_*=[\mbox{dcos}_0(\theta_i-\theta_j)\tilde{x}_j
+\mbox{dsin}_0(\theta_i-\theta_j)\tilde{y}_j]
$ where
\begin{align*}
\mbox{dcos}_0(s) & = \left\{
                \begin{array}{ll}
                  \frac{\cos(s)-1}{s}, & \mathrm{~if~} s \neq 0  \\
                  0,  & \mathrm{~if~} s = 0  \\
                \end{array}
              \right.,\\
\mbox{dsin}_0(s) & = \left\{
                \begin{array}{ll}
                  \frac{\sin(s) }{s}, & \mathrm{~if~} s \neq 0 \\
                  1,  & \mathrm{~if~} s = 0 \\
                \end{array}
              \right..
\end{align*}
Then, the closed-loop system composed of \eqref{eq-robot} and \eqref{control} can be written into
the following compact form
\begin{align}
\dot{{x}} & = p(t)  {y} - k_v \mathcal{L}(\lambda(t)) {x}+\mathcal{L}(\lambda(t),k_v B_* -k_w{y}1_N^{\t})\theta,
\nonumber \\
\dot{{y}}& = -p(t)  {x} +\mathcal{L}(\lambda(t), k_w{x}1_N^{\t})\theta, \nonumber  \\
\dot{\theta} & = p(t) 1_N - k_w \mathcal{L}(\lambda(t))\theta. \label{eq-31-compact}
\end{align}

In the rest of this section, $\Phi$ denotes the family of all solution pairs $([x^{\t},y^{\t},\theta^{\t}]^{\t},\lambda)$ of \eqref{eq-31-compact} such that Assumption \ref{ass-ujc} holds. The closed set
\begin{align}\label{eq-Omega}
\Omega=\left\{ [s1_N^{\t},t1_N^{\t},l1_N^{\t} ]^{\t}\left|~s,t,l\in \mathbb{R}\right.\right\}
\end{align}
is called the consensus subspace. The following result translates the analysis of leaderless consensus into the analysis of an equivalent invariance problem.
\begin{Lemma}\label{lemma-4B3}
Consider the closed-loop system \eqref{eq-31-compact}. Suppose Assumption \ref{ass-ujc} holds, $\Phi^{st}$ is UGB, and $\Omega$ is UGA w.r.t. $\Phi$  and the Euclidean norm. Then, Problem \ref{pro-01} is solved by the distributed controller \eqref{control}.
\end{Lemma}

\begin{Proof}
Since $\Phi^{st}$ is UGB, all solution pairs $([x^{\t},y^{\t},\theta^{\t}]^{\t},\lambda)$ of \eqref{eq-31-compact} can be assumed to be forward complete based on the maximum extension theorem of solutions \cite{Hal}. Since the transformation \eqref{eq-transformation01} is invertible and preserves the Euclidean norm, all solutions of the closed-loop system with respect o the original fixed frame are also forward complete and uniformly global bounded. To solve Problem \ref{pro-01}, it remains to verify equation \eqref{eq-consensus}.

By the Gram-Schmidt orthogonalization theorem \cite{HJ1990},
\begin{align}
& \Vert [x^{\t},y^{\t},\theta^{\t}]^{\t}\Vert _{\Omega} \nonumber\\
&=\sqrt{\left \Vert x-\frac{1_{N}^{\t}{x}}{N}1_{N}\right \Vert^2+\left\Vert y-\frac{1_{N}^{\t}{y}}{N}1_{N}\right \Vert^2+\left \Vert\theta-\frac{1_{N}^{\t}{\theta}}{N}1_{N}\right \Vert ^2}.
\label{eq-gram}
\end{align}
If $\Omega$ is UGA w.r.t. $\Phi$  and the Euclidean norm, then for $i=1,\dots,N$,
\begin{align*}
\lim_{t\rightarrow\infty}\left([\tilde{x}_i(t),
                                   \tilde{y}_i(t),
                                    {\theta}_i(t)]
-
 [\frac{1_{N}^{\tt}{x}(t)}{N},
\frac{1_{N}^{\tt}{y}(t)}{N},
\frac{1_{N}^{\tt}{\theta}(t)}{N} ]
\right) =0
\end{align*}
which in turn implies
\begin{align*}
\lim_{t\rightarrow\infty}\left([\tilde{x}_i(t),
                                   \tilde{y}_i(t),
                                   {\theta}_i(t)]
-
[\tilde{x}_j(t),
\tilde{y}_j(t),
 {\theta}_j(t)]
\right)=0.
\end{align*}
By \eqref{eq-transformation01}, the following can be deduced:
\begin{align*}
&\lim_{t\rightarrow\infty}\left(\begin{bmatrix}
                                    {x}_i(t) \\
                                    {y}_i(t) \\
                                 \end{bmatrix} -
\begin{bmatrix}
 {x}_j(t) \\
 {y}_j(t) \\
\end{bmatrix}
\right)\\
&=\lim_{t\rightarrow\infty}\left(\begin{bmatrix}
             \cos(\theta_i)-\cos(\theta_j) & -\sin(\theta_i)+\sin(\theta_j) \\
             \sin(\theta_i)-\sin(\theta_j) & \cos(\theta_i)-\cos(\theta_j)  \\
\end{bmatrix} \begin{bmatrix}\tilde{x}_i(t) \\
                                    \tilde{y}_i(t) \end{bmatrix} \right.\\
&~~+\left.
\begin{bmatrix}
             \cos(\theta_j) & -\sin(\theta_j) \\
             \sin(\theta_j) & \cos(\theta_j)  \\
\end{bmatrix}
\begin{bmatrix}
 \tilde{x}_i(t)-\tilde{x}_j(t) \\
 \tilde{y}_i(t)-\tilde{y}_j(t) \\
\end{bmatrix}
\right)=0.
\end{align*}
This completes the proof.
\end{Proof}

Lemma \ref{lemma-4B3} has converted the leaderless consensus of the system \eqref{eq-robot} described in Problem \ref{pro-01} into the analysis of uniform global attractivity of the consensus subspace $\Omega$.

\subsection{Uniform Global Attractivity of $\Omega$} \label{sec-4C}

In view of the third equation of \eqref{eq-31-compact}, letting $\hat{\theta}=\theta - (1_N^{\t}\theta /N) 1_N$, one gets
\begin{align}
\dot{\hat{\theta}}   =   - k_w \mathcal{L}(\lambda(t))\hat{\theta}.\label{eq-31-thetahat}
\end{align}
Uniform global attractivity of $\Omega$ w.r.t. $\Phi$ and the Euclidean norm can then be achieved by the following five steps:
\begin{itemize}
  \item  Step 1 shows that $\hat{\theta}=0$ is UGES, which is the key to the rest.
  \item  Step 2 shows that $\Phi^{st}$ is UGB.
  \item  Step 3 shows that $\Omega$ is ULS w.r.t. $\Phi$ and the Euclidean norm.
  \item  Step 4 shows that bounded-output-energy condition \eqref{eq-add-01++} holds.
  \item  Step 5 shows that $\Omega$ is WO w.r.t. $\Phi$.
\end{itemize}
As a consequence of Proposition \ref{corollary-useful}, uniform global attractivity of $\Omega$ w.r.t. $\Phi$ and the Euclidean norm can be obtained. Based on Step 2 and Lemma \ref{lemma-4B3}, the proof of Theorem \ref{Theorem-consensus} is therefore complete.

To make the overall process more digestible, a number of technical lemmas are introduced, the proofs of which may be found in the Appendix to the paper.

\begin{Lemma}\label{lemma-1}
   Consider three functions $\alpha_i:~[s,+\infty) \rightarrow \mathbb{R}_{\geq 0}$, $i=1,2,3$, for some $s\in \mathbb{R}$. Suppose that $\alpha_1$ is locally absolutely continuous, $\alpha_2$ is measurable, and $\alpha_3$ is Lebesgue integrable. If
   \begin{align}\label{eq-18}
\dot{\alpha}_1(t) \leq -{\alpha}_2(t) + \alpha_3(t) (1 + \alpha_1(t))
   \end{align}
for almost all $t\in [s,+\infty)$, then, for all $t\geq s $,
the following inequalities hold:
\begin{align}
  \alpha_1(t) & \leq e^{\int_{s}^{t}\alpha_3(\tau)\mathrm{d}\tau}(1+\alpha_1(s))-1, \label{eq-19}\\
\int_{s}^{+\infty}\alpha_2(\tau)\mathrm{d}\tau    & \leq \beta(s) e^{\beta(s)} +
(1+\beta(s) e^{\beta(s)}) \alpha_1(s) \label{eq-20}
\end{align}
where $0 \leq \beta(s) \triangleq \int_{s}^{+\infty}\alpha_3(\tau)\mathrm{d}\tau< +\infty$.
\end{Lemma}

\begin{Lemma}\label{lemma-4B1}
Let $L_0=I-({1}/{N})1_N1_N^{\t}$. For any $\zeta \in \Lambda$ and any $B=[b_{ij}]\in \mathbb{R}^{N \times N}$, the following hold:
\begin{enumerate}
  \item [1)]  $\Vert L_0\Vert \leq 1$.
  \item [2)] $L_0\mathcal
{L}(\zeta)=  \mathcal
{L}(\zeta)$ and $\mathcal
{L}(\zeta,B)L_0=\mathcal
{L}(\zeta,B)$. Particularly, $L_0^2=L_0$.
  \item [3)] $u^{\t} \mathcal
{L}(\zeta,v 1_N^{\t}) =v^{\t} \mathcal
{L}(\zeta,u 1_N^{\t})$, for any $u,v\in \mathbb{R}^N$.
  \item [4)] $\Vert \mathcal
{L}(\zeta,B)\Vert  \leq \sqrt{N} \Vert \mathcal
{L}(\zeta)\Vert \rho(B)$ where $\rho(B)=(\sum_{i=1}^N\sum_{j=1,j\neq i}^{N}b_{ij}^2)^{{1}/{2}}$.
\end{enumerate}
\end{Lemma}

\begin{Lemma}\label{lemma-4C1}
  Under Assumption \ref{ass-ujc}, there exists $\varepsilon>0$
  such that for any $\lambda \in \Phi^{sw}$, the inequality
  \begin{align}\label{eq-40+}
  u^{\t} \left[\int_{t}^{t+T} \mathcal
{L}(\lambda(\tau))\mathrm{d}\tau \right] u \geq \varepsilon,~\forall t \in \mathbb{R}_{\geq 0}
  \end{align}
holds for any unit vector $u\in \mathbb{R}^N$ satisfying $1_N^{\t} u = 0$ where $T$ is the constant given in Assumption \ref{ass-ujc}.
\end{Lemma}

\subsubsection*{Step 1 -- Uniform global exponential stability of $\hat{\theta}=0$}

  Consider a positive definite and radially unbounded function $W=\Vert  \hat{\theta}\Vert ^2$. Then, the time derivative of $W$ along the trajectories of the system \eqref{eq-31-thetahat}, satisfy that
  \begin{align}\label{eq-38} 
\dot{W}  = - 2 k_w \hat{\theta}^{\t}  \mathcal
{L}(\lambda(t)) \hat{\theta}  \leq 0.
  \end{align}
 So $\hat{\theta}=0$ is UGS \cite{LJ2008}.  Define a virtual output $h_1(\hat{\theta},\zeta)=(2k_w \hat{\theta}^{\t} \mathcal
{L}(\zeta) \hat{\theta})^{{1}/{2}}$. Employing Lemma \ref{lemma-1} with $\alpha_1(t)=W(\hat{\theta}(t))$, $\alpha_2(t)=|h_1(\hat{\theta}(t),\lambda(t))|^2$, and $\alpha_3(t)=0$, we have
\begin{align*}
\int_{s}^{+\infty} |h_1(\hat{\theta}(\tau),\lambda(\tau))|^2\mathrm{d} \tau \leq \Vert \hat{\theta}(s)\Vert ^2,~\forall s \geq 0
\end{align*}
 which results in
\eqref{eq-add-01++}.

Observe that Assumption \ref{ass-zeroingpair} holds with
$\hat{f}=0\cdot 1_N$ and $\hat{h}_1(\hat{\theta},\zeta)= \mathcal
{L}(\zeta) \hat{\theta}$. According to Proposition \ref{prop-1}, the system \eqref{eq-31-thetahat} can then be changed as
\begin{align}
\dot{\hat{\theta}}  = 0,~
\hat{h}_1 = \mathcal
{L}(\lambda(t)) \hat{\theta}. \label{eq-41}
\end{align}
Let $\bar{\hat{\theta}}:~\mathbb{R} \rightarrow \mathbb{R}^N$ be any bounded limiting zeroing-output solution. Then, $\bar{\hat{\theta}}\equiv u_1$ is a constant function and
 $
\lim_{n\rightarrow \infty}  \mathcal
{L}(\lambda_n(t+t_n))u_1= 0
$
for almost all $t$ in $\mathbb{R}$, some $\{t_n\}\subseteq \mathbb{R}_{\geq 0}$  and $\{\lambda_n\} \subseteq \Phi^{sw}$  with $t_n\geq 2n$ for all $n \in \mathbb{N}$.
Notice that $\hat{\theta}=L_0\theta$.
Since $1_N^{\t}\hat{\theta}=0$,
we have $1_N^{\t}u_1=0$ in view of \eqref{eq-12} with
$\chi_{\zeta}=\{v \in \mathbb{R}^N|~1_N^{\t}v=0\}$ for all $\zeta \in \Lambda$.
If $u_1 \neq 0$, according to Lemma \ref{lemma-4C1},
\begin{align*}
\varepsilon  & \leq \lim_{n\rightarrow +\infty} \frac{u_1^{\t}}{\Vert u_1\Vert }  \left(\int_{t_n}^{t_n + T  } \mathcal
{L}(\lambda_n(\tau))\mathrm{d}\tau \right)  \frac{u_1}{\Vert u_1\Vert }  \\
& = \frac{1}{\Vert u_1\Vert ^2}u_1^{\t}\left(\int_{0}^{T  } \lim_{n\rightarrow +\infty}\mathcal
{L}(\lambda_n(\tau+t_n))u_1\mathrm{d}\tau\right)=0
\end{align*}
reaching a contradiction. So $\bar{\hat{\theta}}\equiv 0$, and hence, $\hat{\theta}=0$ is WO w.r.t. $\hat{\Phi}=\{(\hat{\theta}, \lambda)\}$ where $ \hat{\theta}$ is a complete solution of \eqref{eq-31-thetahat} and $\lambda$ is in $\Phi^{sw}$. From Corollary \ref{corollary-family-0}, $\hat{\theta}=0$ is UGAS. Furthermore, since the system \eqref{eq-31-thetahat} is linear,
the so-called scaling invariant property holds. Hence, $\hat{\theta}=0$ is UGES \cite[Lemma 1]{LJ2008}, i.e., there exist $a>0$ and $b>0$ such that
 \begin{align}\label{eq-44}
\Vert \hat{\theta}(t)\Vert  \leq a e^{-b(t-s)}\Vert \hat{\theta}(s)\Vert ,~\forall~t\geq s\geq 0.
 \end{align}
This completes Step 1.

\subsubsection*{Step 2 -- Uniform global boundness of $\Phi^{st}$}
By Lemma \ref{lemma-4B1} and due to $\hat{\theta}=L_0{\theta}$,
 \begin{align}\label{eq-44++}
\Vert \hat{\theta}(t)\Vert  \leq \Vert {\theta}(t)\Vert ,~\forall~t \in \mathbb{R}_{\geq 0}.
 \end{align}
By the mean value theorem, $|\mbox{dcos}_0(s)|\leq 1$ and $|\mbox{dsin}_0(s)|\leq 1$ for any $s\in \mathbb{R}$. Based on the Cauchy inequality, the following holds:
\begin{align}\label{eq-add-002}
&\rho(B_*) \leq  \nonumber\\
&\sqrt{2}( \sum_{i=1}^N\sum_{j=1,j\neq i}^{N}(\mbox{dcos}_0^2(\theta_i-\theta_j) \tilde{x}_j^2 +\mbox{dsin}_0^2(\theta_i-\theta_j)\tilde{y}_j^2))^{\frac{1}{2}} \nonumber\\
&\leq \sqrt{2}( \sum_{i=1}^N (\Vert x\Vert ^2+\Vert y\Vert ^2 ))^{\frac{1}{2}}=\sqrt{2N}(\Vert x\Vert ^2+\Vert y\Vert ^2)^{\frac{1}{2}}.
\end{align}

With a Lyapunov-like function $U=\Vert x(t)\Vert ^2+\Vert y(t)\Vert ^2$, the time derivative of $U$ satisfies
  \begin{align}%
\dot{U}(t)
& = - 2 k_v {x}^{\t}(t) \mathcal
{L}(\lambda(t)) {x}(t) + 2 k_v {x}^{\t} (t) \mathcal{L}(\lambda(t),B_*)\hat{\theta}(t) \nonumber \\
&\leq 2  k_v \sqrt{N} \Vert \mathcal
{L}(\lambda(t))\Vert  \Vert \hat{\theta}(t) \Vert   \Vert x (t)\Vert\rho(B_*) \nonumber\\
&\leq    \tilde{a} e^{-b(t-s)}\Vert \theta(s)\Vert  (1+U(t))\label{eq-51}
  \end{align}
with $\tilde{a}=2\sqrt{2}aNk_v\max_{\zeta\in \Lambda}\Vert \mathcal
{L}(\zeta)\Vert $,
where Lemma \ref{lemma-4B1} and \eqref{eq-44} - \eqref{eq-add-002} were used. Employing Lemma \ref{lemma-1} with $\alpha_1(t)=U(t)$, $\alpha_2(t)=0$, and $\alpha_3(t)=\tilde{a} e^{-b(t-s)}\Vert  {\theta}(s)\Vert $ gives us that
\begin{align}
U(t)   & \leq  F_1^2(x(s),y(s),\theta(s))\label{eq-add-003} 
\end{align}
for any $t \geq s\geq 0$, where $F_1(x,y,\theta)=e^{\frac{\tilde{a} \Vert \theta\Vert }{2b}}(1+x^2+y^2)^{1/2}$.

Notice that $1_N^{\t}\dot{\theta}(t)=Np(t)$. Then, $1_N^{\t}{\theta}(t)=1_N^{\t}{\theta}(s)+N\int_{s}^{t}p(\tau)\mathrm{d}\tau$ for any $t\geq s \geq 0$. This, together with \eqref{eq-30}, \eqref{eq-44},  and \eqref{eq-44++}, gives us that
\begin{align}
& \Vert \theta(t)\Vert \leq\Vert \hat{\theta}(t)\Vert + \left\Vert \frac{1_N^{\t}\theta(t) }{N} 1_N \right\Vert  \nonumber\\
 &\leq a\Vert \hat{\theta}(s)\Vert  + \frac{1}{\sqrt{N}} \left(\vert  1_N^{\t}\theta(s)  \vert  + \left|N\int_{s}^{t}p(\tau)\mathrm{d}\tau\right|\right) \nonumber\\
 & \leq F_2( {\theta}(s))\label{eq-add-003++}
\end{align}
for any $t \geq s\geq 0$, where $F_2(\theta)= (a+1) \Vert {\theta} \Vert  +  {\sqrt{N}}$ $\int_{0}^{T_0}\left|p(\tau)\right|\mathrm{d}\tau
$.

Now \eqref{eq-add-003} and \eqref{eq-add-003++} lead to
 \begin{align*}
\Vert [x^{\t}(t),y^{\t}(t),\theta^{\t}(t)]^{\t}\Vert = \left(U(t)+\Vert {\theta}(t)\Vert ^2\right)^\frac{1}{2} \leq F(s)
\end{align*}
for any $t \geq s\geq 0$ where $F(s) =  (F_1^2(x(s),y(s),\theta(s))+F_2^2(\theta(s)))^{1/2}$.  Thus, $\Phi^{st}$ is UGB. This completes Step 2.

\subsubsection*{Step 3 -- Uniform Lyapunov stability of $\Omega$} Similar to \eqref{eq-add-002},
\begin{align}\max\{\rho(x  1_N^{\t}), \rho(y  1_N^{\t})\}\leq  \sqrt{2N}(\Vert x\Vert ^2+\Vert y\Vert ^2)^{\frac{1}{2}}.\label{eq-50-aaa}\end{align}
 Notice that for any $t\geq s \geq 0$,
\begin{align}
&\max\{\Vert  \mathcal{L}(\lambda(t),k_v B_* -k_w{y}1_N^{\t}) \Vert,\Vert\mathcal{L}(\lambda(t),  k_w {x}(t)1_N^{\t})\Vert\}  \nonumber \\
&\leq  {N}^{\frac{1}{2}} \Vert \mathcal{L}(\lambda(t))\Vert \nonumber\\
&~~~~~\cdot \max\{k_v\rho(B_*)+k_w \rho(y(t) 1_N^{\t}),k_w \rho(x(t) 1_N^{\t}) \} \nonumber \\
&\leq \sqrt{2}N (k_v+k_w)\max_{\zeta \in \Lambda} \Vert \mathcal{L}(\zeta)\Vert  F_1(x(s),y(s),\theta(s))  \label{eq-add-00p}
\end{align}
where Lemma \ref{lemma-4B1} and \eqref{eq-44}, \eqref{eq-add-002}, \eqref{eq-add-003}, \eqref{eq-50-aaa} were used.

Consider a Lyapunov like function $V = \Vert [x^{\t}(t),y^{\t}(t),\theta^{\t}(t)]^{\t}\Vert _{\Omega}^2$. By \eqref{eq-gram}, $V = \Vert L_0x(t)\Vert ^2+\Vert L_0y(t)\Vert ^2+\Vert L_0\theta(t)\Vert ^2$.  For any $r > 0$ and any $t\geq s \geq 0$ with $\Vert [x^{\t}(s),y^{\t}(s),\theta^{\t}(s)]^{\t}\Vert \leq r$, the time derivative of $V$ satisfies that
\begin{align*}
\dot{V}(t)&=2p(t)x^{\t}(t)L_0y(t)-2p(t)y^{\t}(t)L_0x(t)\\
&~-2k_vx^{\t}(t)\mathcal
{L}(\lambda(t))x(t)-2k_w\hat{\theta}^{\t}(t)\mathcal{L}(\lambda(t))\hat{\theta}(t)\\
&~+2(L_0x(t))^{\t} \mathcal{L}(\lambda(t),k_v B_* -k_w{y}1_N^{\t})\hat{\theta}(t)\\
&~+2 (L_0y(t))^{\t}\mathcal{L}(\lambda(t), k_w {x}(t)1_N^{\t})\hat{\theta}(t) \\
&\leq  \bar{\bar{a}}(r)  V^{\frac{1}{2}}(s)e ^{-b(t-s)}(1+V(t))
\end{align*} 
with $\bar{\bar{a}}(r) =2\sqrt{2}aN (k_v+k_w)\max_{\zeta \in \Lambda}\Vert \mathcal{L}(\zeta)\Vert e^{\frac{\tilde{a} r}{2b}}(1+r^2)^{\frac{1}{2}} \geq 0,
$ where Lemma \ref{lemma-4B1} and  \eqref{eq-44}, \eqref{eq-44++}, \eqref{eq-add-00p} were used. Employing Lemma \ref{lemma-1} with $\alpha_1(t)=V(t)$, $\alpha_2(t)=0$, and $\alpha_3(t)= \bar{\bar{a}}(r)  V^{{1}/{2}}(s)  e ^{-b(t-s)}$ gives us that
\begin{align*}
V(t) 
&\leq e^{\frac{\bar{\bar{a}}(r)}{b}V^{\frac{1}{2}}(s)}(1+V(s) )-1
\end{align*}
for any $r > 0$ and any $t\geq s \geq0$ with $\Vert [x^{\t}(s),y^{\t}(s),\theta^{\t}(s)]^{\t}\Vert \leq r$.

Notice that $\lim_{l\rightarrow 0^+}(e^{\bar{\bar{a}}(r)l/b}(1+l^2)-1)^{{1}/{2}}=0$. Then, by continuity, for any $\varepsilon>0$, there exists $\delta = \delta(r,\varepsilon)>0$ such that $0 \leq (e^{\bar{\bar{a}}(r)l/b}(1+l^2)-1)^{{1}/{2}} < \varepsilon$, for any $0\leq l < \delta$. Thus, for any $t\geq s \geq 0$, if $\Vert [x^{\t}(s),y^{\t}(s),\theta^{\t}(s)]^{\t}\Vert _{\Omega}< \delta$ and $\Vert [x^{\t}(\tau),y^{\t}(\tau),\theta^{\t}(\tau)]^{\t}\Vert  \leq r$, for all $\tau\geq s$, then
 $\Vert [x^{\t}(t),y^{\t}(t),\theta^{\t}(t)]^{\t}\Vert _{\Omega} < \varepsilon$, for any $t\geq s\geq 0$. which shows that $\Omega$ is ULS w.r.t. $\Phi$ and the Euclidean norm. This completes Step 3.

\subsubsection*{Step 4 -- Bounded-output-energy condition} Define the virtual output
$h(x,\theta,\zeta) = [(2k_v x^{\t} \mathcal {L}(\zeta) x)^{{1}/{2}},\Vert L_0\theta\Vert  ]^{\t}$.  By \eqref{eq-44}, \eqref{eq-44++}, \eqref{eq-add-003}, and the last inequality of \eqref{eq-51},
\begin{align}
 &\Vert 2 k_v {x}^{\t} (t) \mathcal{L}(\lambda(t),B_*)\hat{\theta}(t) \Vert \leq \nonumber\\
 &~~~~~ \tilde{a}\Vert \theta(s)\Vert  (1+F_1^2(x(s),y(s),\theta(s)))e^{-b(t-s)}. \label{eq-add-002aa+}
\end{align}
It follows that
\begin{align}
&\int_{s}^{t}\Vert h(x(\tau),\theta(\tau),\lambda(\tau))\Vert ^2\mathrm{d} \tau \nonumber\\
&=\int_{s}^{t} 2k_w x^{\t}(\tau) \mathcal {L}(\lambda(\tau))x(\tau) \mathrm{d} \tau + \int_{s}^{t} \Vert L_0\theta(\tau)\Vert ^2 \mathrm{d} \tau \nonumber\\
&\leq U(s) +  \int_{s}^{t} 2 k_v {x}^{\t} (\tau) \mathcal{L}(\lambda(\tau),B_*)\hat{\theta}(\tau)\mathrm{d}\tau  +  \frac{a^2}{2b}\Vert L_0\theta(s)\Vert ^2 \nonumber\\
&\leq \Vert x(s)\Vert ^2+\Vert y(s)\Vert ^2+ \frac{\tilde{a}}{b}\Vert {\theta}(s)\Vert \left(1+ F_1^2(x(s),y(s),\theta(s)) \right) \nonumber\\
&~+  \frac{a^2}{2b}\Vert \theta(s)\Vert ^2 \label{eq-add-009}
\end{align}
for any $t\geq s \geq 0$, where \eqref{eq-44}, \eqref{eq-44++}, \eqref{eq-add-002aa+}, and the first equality of \eqref{eq-51} were used. Letting $t\rightarrow \infty$, the bounded-output-energy condition \eqref{eq-add-01++} holds. This completes Step 4.

\subsubsection*{Step 5 -- Weak observability of $\Omega$}

By zeroing the virtual output $h$ and using Lemma \ref{lemma-4B1}, Assumption \ref{ass-zeroingpair} holds with
$\hat{f}=[p(t)y^{\t},-p(t)x^{\t},p(t)1_N]^{\t}$ and $\hat{h}=h$, where for any $\zeta \in \Lambda$, and $B \in \mathbb{R}^{N \times N}$, $\mathcal{L}(\zeta,B)\theta=\mathcal{L}(\zeta,B)(L_0)\theta$ was used. Then, the system \eqref{eq-31-compact} can be changed as
\begin{align}
\dot{x}=p(t)y,~\dot{y}=-p(t)x,~\dot{\theta}=p(t)1_N. \label{eq-31-changed}
\end{align}
Let $[\bar{x}^{\t},\bar{y}^{\t},\bar{\theta}^{\,\t}]^{\t}:$ $\mathbb{R} \rightarrow \mathbb{R}^{3N}$ be any bounded limiting zeroing-output solution.
Then,  $[\bar{x}^{\t},\bar{y}^{\t},\bar{\theta}^{\,\t}]^{\t}$ satisfies \eqref{eq-13} and \eqref{eq-14} for some $\{t_n\}\subseteq \mathbb{R}_{\geq 0}$  and $\{\lambda_n\} \subseteq \Phi^{sw}$  with $t_n\geq 2n$.
By the compactness of $[0,T_0]$, there exist $0\leq t_0 \leq T_0$ and a subsequence $\{t_{n_k}\}$ of
$\{t_{n}\}$ such that $\lim_{k\rightarrow \infty}(t_{n_k}-\lfloor{t_{n_k}}/{T_0}\rfloor T_0) = t_0$
where $T_0$ is the period of $p(t)$ and $\lfloor{t_{n_k}}/{T_0}\rfloor$ denotes the greatest integer less than or equal to ${t_{n_k}}/{T_0}$.
As a consequence,
 $\lim_{k\rightarrow \infty}p(t+t_{n_k})=p(t+t_0)$ for almost all $t\in \mathbb{R}$.
Thus, $[\bar{x}^{\t}(t),\bar{y}^{\t}(t),\bar{\theta}^{\,\t}(t)]^{\t}$ is a solution of the system \eqref{eq-31-changed} where $p(t)$ is replaced by $p(t+t_0)$.
For simplicity, consider the transformation $[\tilde{\bar{x}}^{\t}(t),\tilde{\bar{y}}^{\t}(t),\tilde{\bar{\theta}}^{\,\t}(t)]^{\t}=[\bar{x}^{\t}(t-t_0),\bar{y}^{\t}(t-t_0),\bar{\theta}^{\,\t}(t-t_0)]^{\t}$
for all $t\in \mathbb{R}$. Then, $[\tilde{\bar{x}}^{\t},\tilde{\bar{y}}^{\t},\tilde{\bar{\theta}}^{\,\t}]^{\t}$ is just a solution of the system \eqref{eq-31-changed}.

Let $I_{0}^m=[mT_0/2, T + mT_0/2)$ and $I_{1}^m=[T + mT_0/2,(m+1)T_0/2)$ for any $m\in \mathbb{Z}$. We then have $I_0^m \cup I_1^m = [mT_0/2, (m+1)T_0/2)$ and $\mathbb{R} = \cup_{m\in \mathbb{Z}}(I_0^m \cup I_1^m)$.
In view of \eqref{eq-62}, due to $p(t)=0$, we have
\begin{align}\label{eq-add-00+2-b}
 \tilde{\bar{x}} (t)=u_m,~\tilde{\bar{y}} (t) =v_m
\end{align}
for all $t\in I_0^m$ and for some constants $u_m,v_m\in \mathbb{R}^{N}$.
According to \eqref{eq-14} and the fact that $\mathcal{L}(\lambda(t))$ is positive semi-definite,
\begin{align}\label{eq-add-00+1}
 &\lim_{k\rightarrow +\infty}\mathcal {L}(\lambda_{n_k}(t-t_0+t_{n_k}) )u_m \nonumber\\
 &=\lim_{k\rightarrow +\infty}(\mathcal {L}(\lambda_{n_k}(t-t_0+t_{n_k}))
                                 {\bar{x}}(t-t_0) =0
\end{align}
for almost all $t\in I_0^m$. By Lemma \ref{lemma-4B1}, this further implies $\lim_{k\rightarrow +\infty}\mathcal {L}(\lambda_{n_k}(t-t_0+t_{n_k}))\hat{u}_m=0$ with $\hat{u}_m=L_0 u_m$. If $\hat{u}_m\neq 0$, according to Lemma \ref{lemma-4C1},
\begin{align*}
&\varepsilon  \leq \lim_{k\rightarrow +\infty} \frac{\hat{u}_m^{\tt} }{\Vert \hat{u}_m\Vert }  \left(\int_{-t_0+t_{n_k}+\frac{mT_0}{2}}^{T-t_0+t_{n_k}+\frac{mT_0}{2}} \mathcal
{L}(\lambda_{n_k}(\tau)) \mathrm{d}\tau \right)  \frac{\hat{u}_m}{\Vert \hat{u}_m\Vert } \\
& = \frac{\hat{u}_m^{\t}}{\Vert \hat{u}_m\Vert ^2} \int_{\frac{mT_0}{2}}^{T+\frac{mT_0}{2}}  \lim_{k\rightarrow +\infty}(\mathcal
{L}_{n_k}\lambda(\tau-t_0+t_{n_k}))\hat{u}_m\mathrm{d}\tau =0
\end{align*}
reaching a contradiction.
Hence, $L_0 u_m=\hat{u}_m=0$ for all $m \in \mathbb{Z}$.
By continuity, we then conclude that
\begin{align}\label{eq-add-00+2-a}
L_0\tilde{\bar{x}}(t)=\hat{u}_m=0,
\end{align}
for all $m \in \mathbb{Z}$ and all $t \in [m{T_0}/{2}, T + m{T_0}/{2}]$. For $t\in I_1^m$, $[(L_0\tilde{\bar{x}}(t))^{\t},(L_0\tilde{\bar{y}}(t))^{\t}]^{\t}$ satisfies
\begin{align*}
L_0\dot{{\tilde{\bar{x}}}}&=(-1)^mc(t-T-\frac{mT_0}{2})L_0{\tilde{\bar{y}}},\nonumber\\
L_0\dot{{\tilde{\bar{y}}}}&=-(-1)^{m}c(t-T-\frac{mT_0}{2})L_0{\tilde{\bar{x}}}.
\end{align*}
Solving this equation, one gets
\begin{align}
[L_0{\tilde{\bar{x}}}(t)]_i&=r_{mi}\sin\left((-1)^m\int_{0}^{t-T-\frac{mT_0}{2}}c(\tau)\mathrm{d}\tau +\psi_{mi}\right),\label{eq-add-00+2-solution1}\\
[L_0{\tilde{\bar{y}}}(t)]_i&=r_{mi}\cos\left((-1)^m\int_{0}^{t-T-\frac{mT_0}{2}}c(\tau)\mathrm{d}\tau +\psi_{mi}\right)\label{eq-add-00+2-solution2}
\end{align}
for any $i=1,\dots,N$ and some $r_{mi}\in \mathbb{R}_{\geq 0}$, $\psi_{mi}\in [0,2\pi)$, where $[L_0{\tilde{\bar{x}}}(t)]_i$ and $[L_0{\tilde{\bar{y}}}(t)]_i$ denote the $i$-th entry of $L_0{\tilde{\bar{x}}}(t)$ and $L_0{\tilde{\bar{y}}}(t)$, respectively.
Observe that from \eqref{eq-add-00+2-a}, $$L_0\tilde{\bar{x}}(T +mT_0/2)=L_0\tilde{\bar{x}}((m+1)T_0/2)=0$$ for all $m \in \mathbb{Z}$.
We claim that $r_{mi}=0$ for
all $m\in \mathbb{Z}$ and all $i=1,\dots,N$. Otherwise, in view of \eqref{eq-add-00+2-solution1},
\begin{align*}
 \sin(\psi_{mi})= \sin\left((-1)^m\int_{0}^{-T+\frac{T_0}{2}}c(\tau)\mathrm{d}\tau +\psi_{mi}\right)=0.
\end{align*}
This gives that
$\psi_{mi}=k_1 \pi$ and $(-1)^m\int_{0}^{-T+{T_0}/{2}}c(\tau)\mathrm{d}\tau +\psi_{mi}=k_2 \pi$ for some $k_1,k_2\in\mathbb{Z}$.
Then,
\begin{align*}
\int_{0}^{-T+\frac{T_0}{2}}c(\tau)\mathrm{d} \tau = (-1)^m(k_2-k_1)\pi,
\end{align*}
reaching a contradiction, since $\int_{0}^{-T + T_0/2}c(\tau)\mathrm{d} \tau \neq k\pi$ for all $k\in \mathbb{Z}$. Thus, $r_{mi}=0$ for
all $m\in \mathbb{Z}$ and all $i=1,\dots,N$, which in turn implies that
 $[(L_0\tilde{\bar{x}}(t))^{\t},(L_0\tilde{\bar{y}}(t))^{\t}]^{\t}=0$  for all  $t\in I_1^m$ and all $m \in \mathbb{Z}$. By \eqref{eq-add-00+2-b},
$$L_0v_m = \lim_{t \rightarrow \left(T + \frac{mT_0}{2}\right)^-}L_0\tilde{\bar{y}}(t)=L_0\tilde{\bar{y}}\left(T + \frac{mT_0}{2}\right)=0.$$ As a result, $[(L_0\tilde{\bar{x}}(t))^{\t},(L_0\tilde{\bar{y}}(t))^{\t}]^{\t}=0$  for all  $t\in I_0^m$ and all $m \in \mathbb{Z}$.
Thus, we can conclude that $[(L_0\tilde{\bar{x}}(t))^{\t},(L_0\tilde{\bar{y}}(t))^{\t}]^{\t}= 0$  for all  $t\in \mathbb{R}$, and hence $[(L_0{\bar{x}}(t))^{\t},(L_0{\bar{y}}(t))^{\t}]^{\t}\equiv 0$. This, together with $L_0\bar{\theta}\equiv 0$, gives us that
\begin{align*}
&\Vert [\bar{x}^{\t}(t),\bar{y}^{\t}(t),\bar{\theta}^{\,\t}(t)]^{\t}\Vert _{\Omega} = \\
& ~~~~(\Vert L_0 {\bar{x}}(t)\Vert ^2+\Vert L_0 {\bar{y}}(t)\Vert ^2+\Vert L_0 {\bar{\theta}}(t)\Vert ^2)^{\frac{1}{2}}=0
\end{align*}
for all $t\in \mathbb{R}$,
which shows that $\Omega$ is WO w.r.t. $\Phi$. This completes Step 5.

It therefore follows that Theorem \ref{Theorem-consensus} holds, as Proposition \ref{corollary-useful} implies that $\Omega$ is UGA w.r.t. to $\Phi$ and the Euclidean norm. This concludes the main result for leaderless consensus without dwell-time conditions.

\begin{figure}
\begin{center}
\scalebox{0.50}{\includegraphics[clip,bb=22 571 513 775]{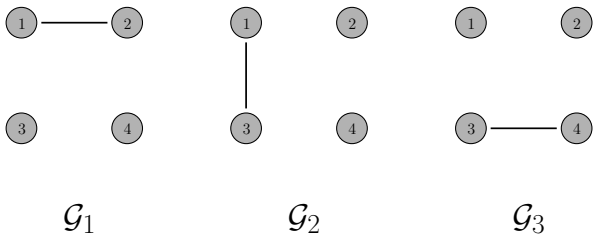}}
\caption{The graphs ${\mathcal {G}}_i$, $i=1,2,3$.}\label{tu-graph}
\end{center}
\end{figure}

\begin{Remark}\label{remark-11-0}
  It is of interest to relate the above attractivity analysis to some other methods in literature.
\begin{itemize}
  \item The closed-loop system \eqref{eq-31-compact} is indeed a switched NLTV system. Therefore, as mentioned in the introduction, most of the early versions of invariance principles \cite{BJ, BM, CWH, GST2008, Hes, MG2011, RSD, SGT2007} for linear/nonlinear time-invariant switched systems cannot be adopted. This challenge is handled here by the new concept of limiting zeroing-output solution as given in Definition \ref{def-3} and the new technique of changing dynamics and output as proposed in Proposition \ref{prop-1}.

  \item Clearly, no dwell-time conditions have been imposed on the switching signal, which makes the result a lot easier to apply in a real world condition. Dwell-time conditions are notoriously hard to enforce in complex switching situations. To the best of the authors' knowledge, there were only invariance principles for linear time-invariant systems presented in \cite{BJ, RSD} without dwell-time constraints, which is based on the linear algebraic techniques. As can be seen from the proof of Theorem \ref{theorem-main}, some techniques from real analysis like the Arzela-Ascoli Lemma  have been employed to address switched NLTV systems. 

 \item It is possible to make use of the well-known Barbalat's Lemma \cite{Kha} (or a generalized version for piecewise continuous signals \cite{SH2012}) to get the convergence of the output signal $h(x(t),\theta(t),\lambda(t))$, in view of \eqref{eq-add-009}. However, by doing so, one may only get the convergence of $\mathcal
{L}(\lambda(t))x(t)$ and $L_0\theta(t)$, but not $L_0x(t)$ and $L_0y(t)$. In contrast, by careful checking the new concept of weak observability as introduced in Definition \ref{Def-5}, we can decide on the convergence of $L_0x(t)$ and $L_0y(t)$. Moveover, Barbalat's Lemma only deals with a single solution, not a family of solutions, demanding additional steps to conclude on uniform attractivity.

 \item Other methods such as the Matrosov-type theorems \cite{ST, TNLT} and the Krasovskii-LaSalle-type theorems \cite{LTM2019} require the attractivity set to be compact and uniformly Lyapunov stable. Since the consensus subspace $\Omega$ given in \eqref{eq-Omega} is merely closed and not compact, neither of these methods are obviously applicable.
\end{itemize}
\end{Remark}

\subsection{Simulation Results} \label{sec-4D}

\begin{figure}
\begin{center}
\scalebox{0.65}{\includegraphics[clip,bb=100 270 495 570]{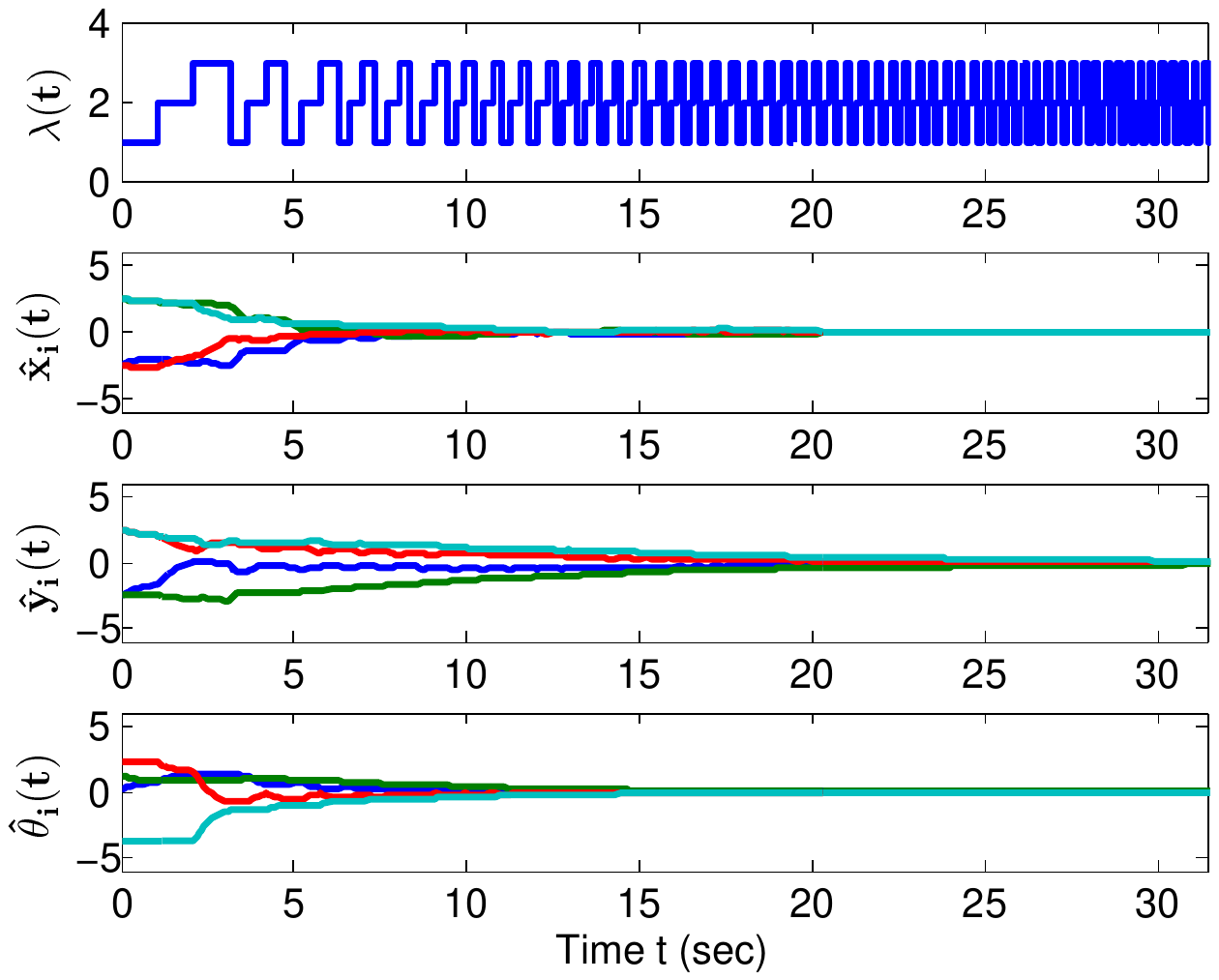}}
\caption{Time responses of the switching signal and the consensus errors.}\label{tu-fig-e}
\end{center}
\end{figure}

An example is provided to illustrate the above control design. Consider a switching graph $\mathcal{G}(\zeta)$, with $\zeta\in \{1,2,3\}$, as shown in Fig. \ref{tu-graph}. The switching signal $\lambda(t)$ is defined as follows:
\begin{align*}
\lambda(t) = \left\{
               \begin{array}{ll}
                 1, & \mbox{~if~} (k+\frac{l}{ k+1})T'\leq t < (k+\frac{l+1/3}{k+1})T', \\
                 2, & \mbox{~if~} (k+\frac{l+1/3}{ k+1})T'\leq t < (k+\frac{l+2/3}{k+1})T',  \\
                 3, & \mbox{~if~} (k+\frac{l+2/3}{ k+1})T'\leq t < (k+\frac{l+1}{k+1})T'   \\
               \end{array}
             \right.
\end{align*}
where $k\in \mathbb{Z}_+$, $l=0,1,\dots,k$, and $T'$ is chosen arbitrarily. It is of interest to see that the proposed switching signal $\lambda(t)$ does not satisfy any dwell-time conditions, since the time slot of two sequenced switching time instants  approaches $0$ as $k\rightarrow+\infty$. This is also illustrated in the first picture  of Fig. \ref{tu-fig-e}.

 Notice that Assumption \ref{ass-ujc} holds with $\tau_a=T'/6$ and $T=T'$. Hence, $\mathcal{G}(\lambda(t))$ is GUJC. Applying Theorem \ref{Theorem-consensus}, a simulation with $k_v=k_w=1$, $c(t)\equiv5$, $T'=\pi$, and $T_0=3T'$ is reported in the last three pictures  of Fig. \ref{tu-fig-e}, in which $\hat{x}_{i}=x_i-(x_1+x_2+x_3+x_4)/4$, $\hat{y}_{i}=y_i-(y_1+y_2+y_3+y_4)/4$, and $\hat{\theta}_i=\theta_i-(\theta_1+\theta_2+\theta_3+\theta_4)/4$ denote the consensus errors. It can be seen that  satisfactory converging behavior is obtained.

\section{Conclusion}\label{sec-conclusion}

A generalization of the LaSalle invariance principle as well as the integral invariance principle for switched NLTV systems was proposed. By introducing a virtual output coming from the derivative of a Lyapunov-like function, with the help of limiting zeroing-output solutions and weak detectability, it is possible to capture the complex limiting behaviors of switched NLTV systems to conclude uniform attractivity with respect to some closed set. Using a consensus problem of mobile robots under switching topology, it was shown how the proposed results can be applied to achieve consensus by design, without requiring uniform Lyapunov stability or imposing unnatural dwell-time conditions. Our future work is directed towards developing the design framework further.

\appendix

\section*{Proofs of Main Theorem \& Corollaries}

\subsection{Proof of Theorem \ref{theorem-main}} \label{sec-05A}

%

%
Suppose $\Omega$ is not UGA w.r.t. $\Phi$  and the Euclidean norm. Then, there exist $\varepsilon_0>0$  and $r_0\geq 0$  such that for each  $T'>0$, there exist $(x,\lambda) \in \Phi$  and a pair $(\bar{s},\bar{t})$ satisfying $\bar{t} \geq \bar{s} + T' \geq \bar{s} \geq \mathrm{Ini}(x)$,
$\Vert x(\bar{s})\Vert \leq r_0$  and $\Vert x(\bar{t})\Vert _{\Omega} \geq \varepsilon_0$.
Since $\Vert x(\bar{s})\Vert  \leq r_0$ and the origin is UGUB, there exist $T_0 > 0$  and $M_0 > 0$ such that $\Vert x(t)\Vert \leq M_0$, for all $ t \geq \bar{s} + T_0$.

Let $T_1=T(\varepsilon_0,M_0)>0$  be the constant given in Assumption \ref{ass-g-convergence}. For each $n\in \mathbb{N}$,  choose $T'=\alpha(n)+T_0+T_1$. Then, there exist $(x_n,\lambda_n)\in \Phi$  and a pair  $(\bar{s}_n,\bar{t}_n)$ satisfying $\bar{t}_n \geq \bar{s}_n + \alpha(n) + T_0 + T_1 \geq \bar{s}_n \geq \mathrm{Ini}(x_n)$, and
\begin{align}\label{eq-A1}
 \Vert x_n(\bar{s}_n)\Vert  \leq r_0 \mbox{~~and~~} \Vert x_n(\bar{t}_n)\Vert _{\Omega} \geq \varepsilon_0.
\end{align}
Let $t'=T_0+T_1+\bar{s}_n-\mathrm{Ini}(x_n) \geq T_1$. Then, we have $\bar{t}_n \geq \alpha(n) + t' + \mathrm{Ini}(x_n)$ and
\begin{align}\label{eq-A2}
\Vert x_n(\tau)\Vert  \leq M_0,~\forall \tau \geq t' + \mathrm{Ini}(x_n) \geq \bar{s}_n + T_0.
\end{align}
According to \eqref{eq-A1} and Assumption \ref{ass-g-convergence}, there exists
\begin{align}\label{eq-A3}
t_n \geq 2n + t' + \mathrm{Ini}(x_n) = 2n + T_0 + T_1 + \bar{s}_n
\end{align}
such that $\Vert x_n(t_n)\Vert _{\Omega} \geq \delta_0 \triangleq \delta(\varepsilon_0,M_0)$ and
the inequality $\int_{t_n-n}^{t_n+n}\Vert h(\tau,x(\tau),\lambda(\tau))\Vert ^2\leq {1}/{n}$ holds. Therefore,
\begin{align*}
0 &\leq \lim_{n\rightarrow \infty} \int_{-n}^{n}\Vert h(\tau+t_n,x_n(\tau+t_n),\lambda_n(\tau+t_n))\Vert ^2\mathrm{d} \tau\\
&= \lim_{n\rightarrow \infty} \int_{t_n-n}^{t_n+n}\Vert h(\tau,x_n(\tau),\lambda_n(\tau))\Vert ^2\mathrm{d} \tau \leq \lim_{n\rightarrow \infty} \frac{1}{n}=0.
\end{align*}

Employing \cite[Lemma A1]{LTM2019} (or its original form \cite[Theorem 1.4 of Chapter 12]{Lan}), there exists a subsequence $\{n_k\}$  of $\{n\}$  such that
\begin{align}\label{eq-A4}
\lim_{k\rightarrow + \infty}h(\tau + t_{n_k},x_{n_k}(\tau+t_{n_k}),\lambda_{n_k}(\tau+t_{n_k})) = 0
\end{align}
for almost all $\tau$ in $\mathbb{R}$. Notice that \eqref{eq-A3} results in $t_{n_k} \geq 2 n_k + T_0 + T_1 + \bar{s}_{n_k} \geq k + T_0 + \bar{s}_{n_k}$, for all $k \in \mathbb{N}$.
Particularly,
\begin{align}\label{eq-A5}
\Vert x_{n_k}(\tau + t_{n_k} )\Vert  \leq M_0, ~\forall k \in \mathbb{N},~\forall -k \leq \tau  \leq k.
\end{align}
Since $f_{\zeta}$, for all $\zeta \in \Lambda$, are almost uniformly bounded, we have
\begin{align*}
&\Vert x_{n_k}(t + t_{n_k} ) - x_{n_k}(s + t_{n_k} )\Vert \\
&=\left \Vert \int_{s + t_{n_k} }^{ t + t_{n_k} }f(\tau,x_{n_k}(\tau),  \lambda_{n_k}(\tau)) \mathrm{d} \tau\right \Vert  \leq M_* |t-s|
\end{align*}
for all $s,t \in [-k,k]$ and some $M_*>0$. For any $\varepsilon >0$, this implies $\Vert x_{n_k}(t + t_{n_k} ) - x_{n_k}(s + t_{n_k} )\Vert < \varepsilon$, for all $s,t \in [-k,k]$ with $|t-s|<  {\varepsilon}/{M_*}$. Thus, $\{x_{n_k}(\cdot + t_{n_k}  ):~[-k,k] \rightarrow \mathbb{R}^{p}\}$ is equi-continuous. Based on \eqref{eq-A5},  $\{x_{n_k}(\cdot + t_{n_k} )\}$ is uniformly bounded. Hence there is a subsequence  $\{x_{n_{k_m}}(\cdot + t_{n_{k_m}} )\}$  of  $\{x_{n_k}(\cdot + t_{n_k} )\}$ converging uniformly to a continuous function $\bar{x}:~\mathbb{R} \rightarrow \mathbb{R}^p $  on every compact subset of $\mathbb{R}$  according to Arzela-Ascoli Lemma \cite[Theorem 3.1 of Chapter 3]{Lan}. By \eqref{eq-A3}, \eqref{eq-A4}, and noticing that $n_{k_m} \geq k_m \geq m$, for all $m \in \mathbb{N}$, the following hold:
\begin{itemize}
  \item $\{t_{n_{k_m}}\} \subseteq \mathbb{R}_{\geq 0}$ and $\{(x_{n_{k_m}},\lambda_{n_{k_m}})\} \subseteq \Phi$.
  \item $t_{n_{k_m}}\geq 2 n_{k_m} + \mathrm{Ini}(x_{n_{k_m}}) \geq 2m + \mathrm{Ini}(x_{n_{k_m}})$.
  \item $\{x_{n_{k_m}}(\cdot + t_{n_{k_m}} )\}$ converges uniformly to $\bar{x}$  on every compact subset of $\mathbb{R}$.
  \item  $\lim_{m\rightarrow + \infty}h(\tau+t_{n_{k_m}},x_{n_{k_m}}(\tau+t_{n_{k_m}}),\lambda_{n_{k_m}}(\tau+t_{n_{k_m}})) = 0$
for almost all $\tau$ in $\mathbb{R}$.
\end{itemize}
Thus, $\bar{x}$  is a limiting zeroing-output solution. By  \eqref{eq-A2} and \eqref{eq-A3}, we have $\Vert \bar{x}(\tau)\Vert  = \lim_{m \rightarrow +\infty}\Vert x_{n_{k_m}}(\tau+t_{n_{k_m}})\Vert \leq M_0 < + \infty$, for all $\tau \in \mathbb{R}$. According to weak observability,
$0 < \delta_0 \leq \lim_{m \rightarrow +\infty}\Vert x_{n_{k_m}}(t_{n_{k_m}})\Vert _{\Omega}=\Vert \bar{x}(0)\Vert _{\Omega}=0$, reaching a contradiction. Therefore, $\Omega$ must be UGA w.r.t. $\Phi$  and the Euclidean norm. This completes the proof.

\subsection{Proof of Corollary \ref{corollary-single}} \label{sec-05B}

Let $\Phi = \{(x,\lambda)\}$. Since $x$ is bounded, $\Phi^{st}$ is UGB (and hence UGUB) w.r.t. $\Phi$. In view of Theorem \ref{theorem-main}, it remains to show that Assumption \ref{ass-3} implies Assumption \ref{ass-g-convergence} in such a case.

Actually, under Assumption \ref{ass-3}, for each  $n \in \mathbb{N}$, there exists $T_n\in \mathbb{N}$ such that
\begin{align*}
\int_{t}^{t+T_0}\Vert h(\tau,x(\tau),\lambda(\tau))\Vert ^2 \mbox {d} \tau \leq \frac{T_0}{2n(n+T_0)},~\forall t \geq T_n.
\end{align*}
Let $\varepsilon \geq 0$  and $r>0$  be any constants. Choose $T=T_0$, $\delta = \varepsilon$, and $\alpha(n)=T_n+2n$, for all $n \in \mathbb{N}$.
Let $t\geq T$ be any constant. Assume that there exists $s' \geq \alpha(n)+t+\mathrm{Ini}(x)$ such that $\Vert x(s')\Vert _{\Omega}\geq\varepsilon$. Let $s=s'\geq\alpha(n)+t+\mathrm{Ini}(x)\geq 2n+t+\mathrm{Ini}(x)$. Then, we have $\Vert x(s)\Vert _{\Omega}\geq\varepsilon=\delta$. Furthermore, let $\lceil{n}/{T_0}\rceil$ be the minimum positive integer that is larger than or equal to ${n}/{T_0}$. So $n \leq \lceil {n}/{T_0}\rceil T_0 < ({n}/{T_0}+1)T_0 = n+T_0$. Hence,
\begin{align*}
s - \left\lceil\frac{n}{T_0}\right\rceil T_0 &\geq \alpha(n) - \left\lceil\frac{n}{T_0}\right\rceil T_0 + t + \mathrm{Ini}(x) \\
&>
T_n + 2n - n - T_0 + T_0 \geq T_n.
\end{align*}
This leads to following inequality
\begin{align*}
&\int_{s-n}^{s+n} \Vert h(\tau,x(\tau),\lambda(\tau))\Vert ^2 \mathrm{d} \tau \\
&\leq
\int_{s-\lceil\frac{n}{T_0}\rceil T_0}^{s+\lceil\frac{n}{T_0}\rceil T_0} \Vert h(\tau,x(\tau),\lambda(\tau))\Vert ^2 \mathrm{d} \tau \\
&\leq \sum_{i=-\lceil\frac{n}{T_0}\rceil}^{\lceil\frac{n}{T_0}\rceil-1}
\int_{s+iT_0}^{s+(i+1) T_0} \Vert h(\tau,x(\tau),\lambda(\tau))\Vert ^2 \mathrm{d} \tau \\
&\leq 2 \lceil\frac{n}{T_0}\rceil \max_{-\lceil\frac{n}{T_0}\rceil \leq i < \lceil\frac{n}{T_0}\rceil} \left\{\int_{s+iT_0}^{s+iT_0+T_0} \Vert h(\tau,x(\tau),\lambda(\tau))\Vert ^2 \mathrm{d} \tau\right\}\\
&\leq \frac{2 \lceil\frac{n}{T_0}\rceil T_0}{2n(n+T_0)} \leq \frac{1}{n}.
\end{align*}
Thus, Assumption \ref{ass-g-convergence} holds. This completes the proof.

\subsection{Proof of Corollary \ref{corollary-family}} \label{sec-05C}

In view of Theorem \ref{theorem-main}, it remains to show that Assumption \ref{ass-4}, together with the fact that $\Omega$ is UULS w.r.t. $\Phi$ and the Euclidean norm, implies Assumption \ref{ass-g-convergence}.

Let $\varepsilon>0$  and $r>0$ be any constants.  For each $n \in \mathbb{N}$, let
$\varepsilon_n=\min\left\{{1}/{r},{1}/{4n^2}\right\}$. Then, $0<\varepsilon_n<1$. According to Assumption \ref{ass-4}, there exist $T_n\triangleq T(\varepsilon_n)>0$  and  $M_n\triangleq M(\varepsilon_n)>0$ such that
\begin{align}\label{eq-a08}
\int_{s}^{t} \Vert h(\tau,x(\tau),\lambda(\tau))\Vert ^2 \mathrm{d} \tau \leq M_n+\varepsilon_n(t-s)
\end{align}
for any $(x,\lambda) \in \Phi$  and any $(s,t)$ with $\mathrm{Ini}(x)+T_n \leq s \leq t$ and $\Vert x(\tau)\Vert \leq {1}/{\varepsilon_n}$, $\forall s\leq \tau \leq t$.
Without lose of generality, we may assume that $T_n$ and $M_n$ are both positive integers by choosing larger constants.

Choose  $\alpha(n)=4n^2M_n+2nT_n$, $\forall n \in \mathbb{N}$. Let $T_0\triangleq T(\varepsilon,r)>0$ and $\delta_0\triangleq \delta(\varepsilon,r)>0$ be the constants given in the definition of  UULS (see Definition \ref{def-02}). For any $t'\geq T_0$, let $(x,\lambda) \in \Phi$ with
\begin{align}\label{eq-a06}
  \Vert x(\tau)\Vert _{\Omega} \leq r \leq \frac{1}{\varepsilon_n},~\forall \tau \geq t'+\mathrm{Ini}(x),
\end{align}
and $\Vert x(s')\Vert _{\Omega}\geq \varepsilon$ for some $s' \geq \alpha(n) +t' + \mathrm{Ini}(x)$. We claim that
\begin{align}\label{eq-a07}
  \Vert x(\tau)\Vert _{\Omega} \geq \delta,~\forall t'+\mathrm{Ini}(x) \leq \tau \leq \alpha(n)+t'+\mathrm{Ini}(x).
\end{align}
Otherwise, there exists $s''$ satisfying $$T_0+\mathrm{Ini}(x) \leq t'+\mathrm{Ini}(x) \leq s''\leq\alpha(n)+t'+\mathrm{Ini}(x) \leq s'$$
such that $\Vert x(s'')\Vert _{\Omega}< \delta$. Since $\Omega$ is UULS w.r.t. $\Phi$ and the Euclidean norm, it holds that $\Vert x(s')\Vert _{\Omega}<\varepsilon$ in view of \eqref{eq-a06}, reaching a contradiction. Thus, \eqref{eq-a07} holds.

Let $s=(2i_n+1)n+t'+\mathrm{Ini}(x)$ for some $T_n \leq i_n \leq 2nM_n+T_n-1$ satisfying that
\begin{align*}
&\int_{2i_nn}^{2(i_n+1)n}\Vert y(\tau)\Vert   ^2 \mathrm{d} \tau = \\
&~~~~~~~\min_{T_n\leq i \leq 2nM_n+T_n-1}\int^{2(i+1)n}_{2i n}\Vert y(\tau)\Vert   ^2 \mathrm{d} \tau,
\end{align*}
where $y(\tau)\triangleq h(\tau+t'+\mathrm{Ini}(x),x(\tau+t'+\mathrm{Ini}(x)),\lambda(\tau+t'+\mathrm{Ini}(x)))$.
Then, we have
$s\geq 2n+t'+\mathrm{Ini}(x)$, and by \eqref{eq-a08} and \eqref{eq-a06},
\begin{align*}
&\int_{s-n}^{s+n}\Vert h(\tau,x(\tau),\lambda(\tau))\Vert ^2 \mathrm{d} \tau \nonumber\\
&=\min_{T_n\leq i \leq 2nM_n+T_n-1}\int^{2(i+1)n}_{2i n}\Vert y(\tau)\Vert   ^2 \mathrm{d} \tau \nonumber \\
&\leq \frac{1}{2nM_n}\int^{2(2nM_n+T_n)n}_{2T_nn} \Vert y(\tau)\Vert   ^2 \mathrm{d} \tau \nonumber \\
& \leq \frac{M_n}{2nM_n}+\frac{ 4 n^2M_n\varepsilon_n}{2nM_n}\leq\frac{1}{2n}+\frac{1}{2n} = \frac{1}{n}.
\end{align*}
Notice that
$$t' +\mathrm{Ini}(x) \leq  (2 i_n + 1) n + t' + \mathrm{Ini}(x) \leq \alpha(n) + t' + \mathrm{Ini}(x).$$
Then, by \eqref{eq-a07}, we have $\Vert x(s)\Vert _{\Omega} \geq \delta$. Thus, Assumption \ref{ass-g-convergence} holds. This completes the proof.

\section*{Proofs of Technical Lemmas}

\subsection{Proof of Lemma \ref{lemma-1}} \label{appendix-E}

Denote
$$\alpha(t)=1+\alpha_1(t) - e^{\int_{s}^{t}\alpha_3(\tau) \mathrm{d}\tau}(1+\alpha_1({s})),~\forall t \geq s.$$ Then, it can be verified that $\alpha(\cdot)$ is locally absolutely continuous. In view of \eqref{eq-18} and the fact that $\alpha_2(\cdot) \geq 0$  and $\alpha_3(\cdot)$ is Lebesgue integrable, we have
\begin{align*}
 &\dot{\alpha}(t) = \dot{\alpha}_1(t) - \alpha_3(t) e^{\int_{{s}}^{t}\alpha_3(\tau) \mathrm{d}\tau}(1+\alpha_3({s})) \\
 &\leq-{\alpha}_2(t) + \alpha_3(t) (1 + \alpha_1(t)) - \alpha_3(t) e^{\int_{{s}}^{t}\alpha_3(\tau)  \mathrm{d}\tau}(1+\alpha_3({s})) \\
 &\leq \alpha_3(t) \alpha(t)
\end{align*}
for almost all $t\geq s$. This further implies that
\begin{align*}
\frac{d}{\mathrm{d} t}\left[ e^{\int_{{s}}^{t}\alpha_3(\tau)  \mathrm{d}\tau} \alpha(t)\right] =
e^{\int_{{s}}^{t}\alpha_3(\tau) \mathrm{d}\tau} \left(\dot{\alpha}(t)-\alpha_3(t)\alpha(t)\right)\leq 0
\end{align*}
for almost all $t\geq s$.
Thus, $e^{\int_{{s}}^{t}\alpha_3(\tau)  \mathrm{d}\tau} \alpha(t)$ is non-increasing, and hence for all $t \geq {s}$,
\begin{align*}
\alpha(t) \leq e^{\int_{{s}}^{t}\alpha_3(\tau)  \mathrm{d}\tau} \alpha(t) \leq e^{\int_{{s}}^{{s}}\alpha_3(\tau)  \mathrm{d}\tau} \alpha({s}) = \alpha({s}) =0.
\end{align*}
Therefore, $1+\alpha_1(t) - e^{\int_{{s}}^{t}\alpha_3(\tau) \mathrm{d}\tau}(1+\alpha_1({s}))\leq0$ for all $t \geq {s}$, which results in \eqref{eq-19}.

Now integrating both sides of \eqref{eq-18} and by using \eqref{eq-19}, one gets
\begin{align*}
 \int_{{s}}^{t} \alpha_2(\tau) \mathrm{d} \tau
 & \leq \alpha_1({s}) - \alpha_1(t)
 + \max_{{s} \leq \tau \leq t}\{1+\alpha_1(\tau)\} \int_{{s}}^{t} \alpha_3(\tau) \mathrm{d} \tau\\
 &\leq \alpha_1({s}) + (1+\alpha_1({s})) e^{\int_{{s}}^{t}\alpha_3(\tau)\mathrm{d}\tau}\int_{{s}}^{t}\alpha_3(\tau)\mathrm{d}\tau
\\
&\leq \alpha_1({s}) + r(s) e^{r(s)} (1+\alpha_1({s}))
\end{align*}
for all $t\geq s$.
By taking the limit, \eqref{eq-20} holds. This completes the proof.

\subsection{Proof of Lemma \ref{lemma-4B1}} \label{appendix-F}

1) Since $L_0$ is symmetric with all eigenvalues being $0$ or $1$, $\Vert L_0\Vert \leq1$.

2) This property can be directly verified by noticing that $1_N^T\mathcal
{L}(\zeta)=0$ and $\mathcal
{L}(\zeta,B)1_N=0$.

3) Let $u=[u_1,\dots,u_N]^{\t}$ and $v=[v_1,\dots,v_N]^{\t}$. Notice that the $i$-th entry of the low vector $u^{\t} \mathcal
{L}(\zeta,v 1_N^{\t})$ is
\begin{align*}
&-\sum_{j=1,j\neq i}^{N}u_j a_{ji}(\zeta)v_{j} + u_i\sum_{j=1}^{N}a_{ji}(\zeta)v_{i} \\
&=-\sum_{j=1,j\neq i}^{N}v_j a_{ji}(\zeta)u_{j} + v_{i}\sum_{j=1}^{N}a_{ji}(\zeta) u_i
\end{align*}
which is exactly the $i$-th entry of the low vector $v^{\t} \mathcal
{L}(\zeta,u 1_N^{\t})$.

4) Recall that for any $C=[c_{ij}]\in \mathbb{R}^{N\times N}$, we have
$$\max_{1\leq i,j\leq N}|c_{ij}|\leq\Vert C\Vert \leq(\sum_{i=1}^N\sum_{j=1}^{N}c_{ij}^2)^{\frac{1}{2}}$$ by the equivalence of matrix norms \cite[Page 314]{HJ1990}.
Then, together with the Cauchy inequality, we have
\begin{align*}
& \Vert \mathcal
{L}(\zeta,B)\Vert ^2
\leq \sum_{i=1}^N\sum_{j=1,j\neq i}^N a_{ij}^2(\zeta)b_{ij}^2 + \sum_{i=1}^{N} (\sum_{j=1,j\neq i}^N a_{ij}(\zeta)b_{ij} )^2 \\
&\leq\Vert \mathcal
{L}(\zeta)\Vert ^2 \sum_{i=1}^N\sum_{j=1,j\neq i}^N b_{ij}^2  + \sum_{i=1}^N (\sum_{j=1,j\neq i}^N a_{ij}^2(\zeta))(\sum_{j=1,j\neq i}^N b_{ij}^2)\\
&\leq N \Vert \mathcal
{L}(\zeta)\Vert ^2 \sum_{i=1}^N\sum_{j=1,j\neq i}^N b_{ij}^2
\end{align*}
which gives that $\Vert \mathcal
{L}(\zeta,B)\Vert  \leq \sqrt{N} \Vert \mathcal
{L}(\zeta)\Vert \rho(B)$. This completes the proof.

\subsection{Proof of Lemma \ref{lemma-4C1}} \label{appendix-H}


Let $(\tau_a,T)$ be the pair as in Assumption \ref{ass-ujc}. Notice that for any $t \in \mathbb{R}_{\geq 0}$ and any vector $u\in\mathbb{R}^N$,
\begin{align*}
u^{\t} \left[\int_{t}^{t+T} \mathcal
{L}(\lambda(\tau))\mathrm{d}\tau\right] u\geq \tau_a u^{\t} \left[\sum_{\zeta \in \lambda_{\tau_a}[t,t+T)}\mathcal {L}(\zeta)\right] u.
\end{align*}
It can be verified that $\sum_{\zeta \in \lambda_{\tau_a}[t,t+T)} \mathcal
{L}(\zeta)$ is a Lapalacian matrix of a connected graph $\mathcal{G}^{\tau_a}_{\lambda}([t,t+T))$.  Denote $\mathbb{S}$ the cluster of the subset $S$ of $\Lambda$ in which
$(\mathcal{V},\cup_{\zeta \in {S}}\mathcal{E}(\zeta))$ is a connected graph. Notice that $\mathbb{S}$ is a finite set. Let
$\varepsilon'=\min_{S\in\mathbb{S}}\left\{\sigma_{\min}\left(\sum_{\zeta \in S}\mathcal
{L}(\zeta)\right)\right\}>0$, where for any symmetric matrix $A$, $\sigma_{\min}(A)$ denotes its minimum positive eigenvalue. According to \cite[Lemma 3.3]{RB2005},
the inequality $u^{\t} \left[\sum_{\zeta \in \lambda_{\tau_a}[t,t+T)}\mathcal{L}(\zeta)\right] u \geq \varepsilon' \Vert u\Vert ^2$ holds for all $u \in \mathbb{R}^N$ with $1_N^{\t} u=0$.
Therefore, \eqref{eq-40+} holds with $\varepsilon=\tau_a\varepsilon'$. This completes the proof.

%
%

\begin{IEEEbiography}[]{Ti-Chung Lee} received the M.S. degree in mathematics and the Ph.D. degree in electrical engineering from the National Tsing Hua University, Hsinchu, Taiwan, in 1990 and 1995, respectively. In August 1997, he joined the Minghsin University of Science and Technology at Hsinchu as an Assistant Professor of Electrical Engineering, and since 2005 he has been a Professor. His main research interests are stability theory, tracking control of nonholonomic systems, and robot control.
\end{IEEEbiography}

\vspace{-0.8cm}

\begin{IEEEbiography}[]{Ying Tan} is a full professor at the Department of Mechanical Engineering (DEEE) at The University of Melbourne, Australia. She received her Bachelor’s degree from Tianjin University, China, in 1995, and her PhD from the National University of Singapore in 2002. She joined McMaster University in 2002 as a postdoctoral fellow in the Department of Chemical Engineering. Since 2004, she has been with the University of Melbourne. She was awarded an Australian Postdoctoral Fellow (2006-2008) and a Future Fellow (2009-2013) by the Australian Research Council. Her research interests are in data-driven optimization, sampled-data distributed systems, rehabilitation robotic systems, and human-robot interfaces.
\end{IEEEbiography}

\vspace{-0.8cm}

\begin{IEEEbiography}[]{Youfeng Su} is a professor in the College of Mathematics and Computer Science at Fuzhou University, Fuzhou, China. He received the B.S. degree in 2005 and the M.S. degree in 2008, both from East China Normal University, Shanghai, China, and the Ph.D. degree in 2012 from The Chinese University of Hong Kong, Hong Kong, China.
 From May 2012 to Jun. 2013, he was a Postdoctoral Fellow at The Chinese University of Hong Kong. Since 2013, he has been with Fuzhou University. His research interests include cooperative control, multi-agent systems, output regulation, and switched/bybrid systems.
\end{IEEEbiography}

\vspace{-0.8cm}

\begin{IEEEbiography}[]{Iven Mareels} is the Director of IBM Research in Australia and honorary
Professor of Electrical and Electronic Engineering at the University of
Melbourne. He is a leading expert in the area of large scale systems, adaptive
control and extremum seeking. He is a Fellow of the IEEE (USA); Fellow of
IFAC (Austria) (he was Vice President of the International Federation of
Automatic Control (IFAC) and Chair of the IFAC Technical Board
(2008-2014)); Fellow of the Australian Academy of Technology and
Engineering (Australia) (and a member of the board of this Academy), and a
(Foreign) Fellow of the Flemish Royal Belgian Academy of Sciences and
Humanities (Belgium). He has co-authored 5 books, and in excess of 150
journal papers and book chapters, and more than 250 conference papers. He
holds 32 international patents.
\end{IEEEbiography}

\vfill

\end{document}